\documentclass{egpubl}
\usepackage{eg2024}
\ConferencePaper
\CGFccby

\usepackage[T1]{fontenc}
\usepackage{dfadobe}  

\biberVersion
\BibtexOrBiblatex
\usepackage[backend=biber,bibstyle=EG,citestyle=alphabetic,backref=true]{biblatex} 
\electronicVersion
\PrintedOrElectronic

\ifpdf \usepackage[pdftex]{graphicx} \pdfcompresslevel=9
\else \usepackage[dvips]{graphicx} \fi

\usepackage{egweblnk} 

\usepackage{amsmath}
\usepackage{comment}
\usepackage{kbordermatrix}
\usepackage{multirow,bigdelim}
\usepackage{cleveref}
\usepackage{algpseudocode}
\usepackage{algorithm}
\usepackage{bm}
\usepackage{bbold}

\usepackage{dsfont}

\usepackage{array}
\usepackage{wrapfig}

\usepackage{csvsimple}
\usepackage{adjustbox}
\usepackage[percent]{overpic}

\newcommand{\nhist}{n_{\mathrm{hist}}}
\newcommand{\nconn}{n_{\mathrm{conn}}}
\newcommand{\pgeo}{p_{\mathrm{geo}}}
\newcommand{\bbv}{{\mathbf{V}}}
\newcommand{\bbu}{{\mathbf{U}}}
\newcommand{\bbt}{{\mathbf{T}}}
\newcommand{\bbd}{{\mathbf{d}}}
\newcommand{\bbD}{{\mathbf{D}}}
\newcommand{\bbs}{{\mathbf{s}}}

\newcommand{\bbn}{{\mathbf{n}}}
\newcommand{\bbsigma}{{\Sigma}}
\newcommand{\bbc}{{\mathbf{c}}}
\newcommand{\bbq}{{{q}}}
\newcommand{\bbf}{{\mathbf{f}}}
\newcommand{\bbF}{{\mathbf{F}}}
\newcommand{\bbQ}{{\mathbf{Q}}}
\newcommand{\bbA}{{\mathbf{A}}}

\newcommand{\Loss}{\mathcal{L}}

\usepackage{color}
\usepackage{soul}
\usepackage{breqn}
\usepackage{booktabs}
\usepackage{tabularx}
\usepackage{rotating}

\definecolor{purple}{rgb}{0.99,0.2,0.72}
\definecolor{blue}{rgb}{0, 0.2, 0.8}
\definecolor{orange}{rgb}{0.6, 0.6, 0}
\definecolor{red}{rgb}{0.8, 0.2, 0.2}
\definecolor{magenta}{rgb}{0.5, 0.0, 1.0}
\definecolor{black}{rgb}{0.0, 0.0, 0.0}
\definecolor{cyan}{rgb}{0, 0.65, 0.65}
\definecolor{olive}{rgb}{0.2, 0.6, 0.5}
\usepackage{xcolor}

\newif\ifdraft
\drafttrue
\ifdraft
\newcommand{\wtc}[1]{{\color{magenta}[\textbf{Tuanfeng:} \textit{#1}]}}
\newcommand{\dcc}[1]{{\color{orange}[\textbf{Duygu:} \textit{#1}]}}
\newcommand{\plc}[1]{{\color{blue}[\textbf{Peizhuo:} \textit{#1}]}}
\newcommand{\OSHc}[1]{{\color{purple}[\textbf{Olga:} \textit{#1}]}}
\newcommand{\tkc}[1]{{\color{olive}[\textbf{Timur:} \textit{#1}]}}

\else
\newcommand{\wtc}[1]{}
\newcommand{\dcc}[1]{}
\newcommand{\plc}[1]{}
\newcommand{\OSHc}[1]{}
\newcommand{\tkc}[1]{}

\fi

\usepackage{pgfplots}
\pgfplotsset{compat=newest}
\usepgfplotslibrary{groupplots}
\usepgfplotslibrary{dateplot}
\usepackage[bottom]{footmisc}
\usepackage{lineno}

\title{Neural Garment Dynamics via Manifold-Aware Transformers}

\author[Li, P.\ et al.] 
{\parbox{\textwidth}{\centering Peizhuo Li$^{1}$\orcid{0000-0001-9309-9967}, Tuanfeng Y. Wang$^{2}$\orcid{0000-0002-8180-4988}, Timur Levent Kesdogan$^{1}$\orcid{0009-0006-5839-9677}, Duygu Ceylan$^{2}$\orcid{0000-0001-6530-4556}, Olga Sorkine-Hornung$^{1}$\orcid{0000-0002-8089-3974}
        }
        \\
{\parbox{\textwidth}{\centering $^1$ETH Zurich, Switzerland $\quad \quad \quad$
         $^2$Adobe Research, United Kingdom
       }
}
}

\begin{document}

\teaser{
\vspace{-0.5cm}
    \includegraphics[width=0.9\linewidth]{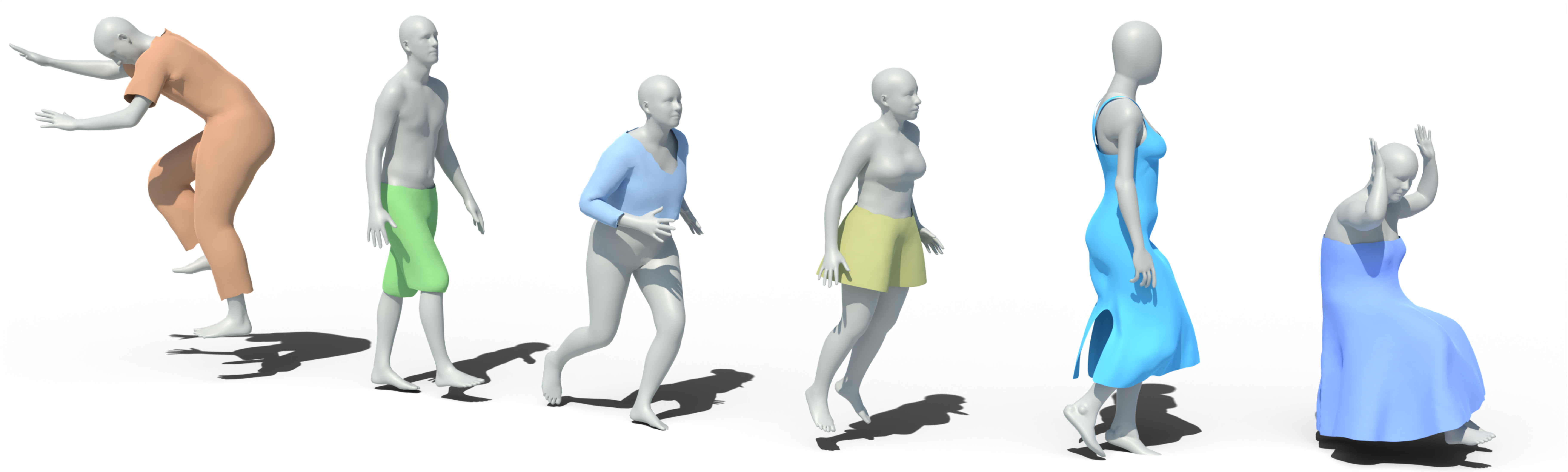}
 \centering
  \caption{We propose a neural garment dynamics inference network powered by manifold-aware transformers. Our approach can be directly applied to unseen
garments, bodies, as well as motions that were not included in the training data.}
\label{fig:teaser}
}
\maketitle

\begin{abstract}
    Data driven and learning based solutions for modeling dynamic garments have significantly advanced, especially in the context of digital humans. However, existing approaches often focus on modeling garments with respect to a fixed parametric human body model and are limited to garment geometries that were seen during training. In this work, we take a different approach and model the dynamics of a garment by exploiting its local interactions with the underlying human body. Specifically, as the body moves, we detect local garment-body collisions, which drive the deformation of the garment. At the core of our approach is a mesh-agnostic garment representation and a manifold-aware transformer network design, which together enable our method to generalize to unseen garment and body geometries. We evaluate our approach on a wide variety of garment types and motion sequences and provide competitive qualitative and quantitative results with respect to the state of the art. 
    
\end{abstract}

\section{Introduction}

Modeling the dynamics of garments as they interact with an underlying collider, such as a moving human body, is a core component for many graphics applications, e.g., animation~\cite{wang2018learning}, virtual try-on~\cite{santesteban2021self}, video editing~\cite{yang2016detailed}, etc. 
There are two main directions to tackle this problem, i.e., physically-based simulation~\cite{nealen2006physically,narain2012adaptive}, and learning-based neural approaches~\cite{patel2020tailornet,bertiche2021pbns}.
Physically-based simulation provides a generic framework to produce plausible and accurate geometric details with realistic motion dynamics. However, acquiring and setting the physical parameters used in the simulation is not easy and often the simulation process is sensitive to initial conditions~\cite{zhang2021deep}. 
To address such issues, learning-based approaches are getting popular in recent years. A typical learning-based workflow~\cite{wang2018learning,pan2022predicting} models the garment dynamics as a mapping from the encoded garment and collider (i.e., body) motion features to a latent code representing the garment shape in the next frame. The compact latent representation enables efficient inference and acts as a regularizer. However, such global approaches are difficult to generalize to unseen garment and collider geometries during training.

In this work, we introduce a novel and generalizable learning-based framework for predicting the garment dynamics by modeling the local interaction between the garment and the underlying collider.
Specifically, we represent the garment deformation with a continuous deformation field~\cite{sorkine2004laplacian,lipman2004differential} where we treat each face of the garment geometry as a sample of this field. We define a set of \emph{garment and interaction features} for each face to encode the state of the garment relative to the underlying body as the body moves from the current to the next frame. Such features inherently encode how each local patch on the garment geometry interacts with the underlying body. We accumulate such features over the past few frames to encode the dynamic behavior and augment them with global context (i.e., the global velocity of the garment). Finally, we train a neural network to predict the deformation gradient of each garment face given this local and global context. Given per-face predictions, we perform a Poisson solve to obtain the final garment geometry in the next frame. Our network runs in an auto-regressive manner by utilizing its past predictions when computing the garment and interaction features that are provided as input to the network in future frames.

Central to our approach is a transformer-based network architecture~\cite{vaswani2017attention} that predicts the deformation of a target point (e.g., the centroid of a face) when given tokens that represent the aforementioned features sampled on a random set of points on the garment surface. The transformer architecture is capable of modeling long-range correlations between how different parts of the garment deform. While spatial proximity is a strong cue for similar deformation behavior, in certain cases, spatially close garment parts can have very different dynamic behavior. For example, imagine two sides of a skirt with a cut (see \Cref{fig:dress-cut}) where nearby points on two sides of the cut behave differently. In order to handle such challenging cases, we empower our transformer to be \emph{manifold aware}. Specifically, we utilize the geodesic distance matrix obtained from the rest state of the garment as part of the attention weights. This encourages the predicted deformation to preserve the geodesic distance in the resulting garment geometry.

Our approach generalizes to a variety of garment types and geometries. We evaluate our method on garment and body types and motion sequences unseen during training. Our approach produces plausible garment geometry with vivid dynamics and performs competitively with respect to the state-of-the-art learning-based approaches. In summary, our main contributions are:
\begin{itemize}
    \item We present a generalizable learning-based approach that predicts plausible garment dynamics for unseen garment and body types.
    \item We present a novel \emph{manifold-aware} transformer architecture that incorporates both spatial and topological information and is agnostic to the underlying meshing density, and can generalize to garments with unseen local connectivity changes such as cuts.
\end{itemize}

\section{Related Work}

\subsection{Physics-based Simulation} 
Modeling the dynamics of the garment w.r.t. the underlying collider motion has been studied in computer graphics for more than 30 years~\cite{terzopoulos1987elastically,moore1988collision}. Physics-based methods~\cite{muller2008hierarchical,muller2007position,narain2012adaptive} tend to model the garment dynamics with real-world physics based on material properties and deform them according to laws of physics using time integration and collision response. The focus of this community includes material modeling~\cite{bhat2003estimating,miguel2012data}, mechanical modeling~\cite{choi2005stable,volino2009simple}, as well as collision modeling~\cite{harmon2009asynchronous,li2020codimensional}, and more recently, converting the whole pipeline to a differentiable setup~\cite{li2022diffcloth,liang2019differentiable} for inverse problems. Physics-based workflows often produce high-quality dynamics but suffer from high computational costs and the tediousness of tuning the material property for a desired effect.

\begin{figure*}[t!]
    \newcommand{\infe}{\color[rgb]{0.99,0.72,0.18}}
    \newcommand{\embedding}{\color[rgb]{0.97,0.74,0.84}}
    \newcommand{\outfe}{\color[rgb]{0.07,0.69,0.15}}
    \centering
    \includegraphics[width=\linewidth]{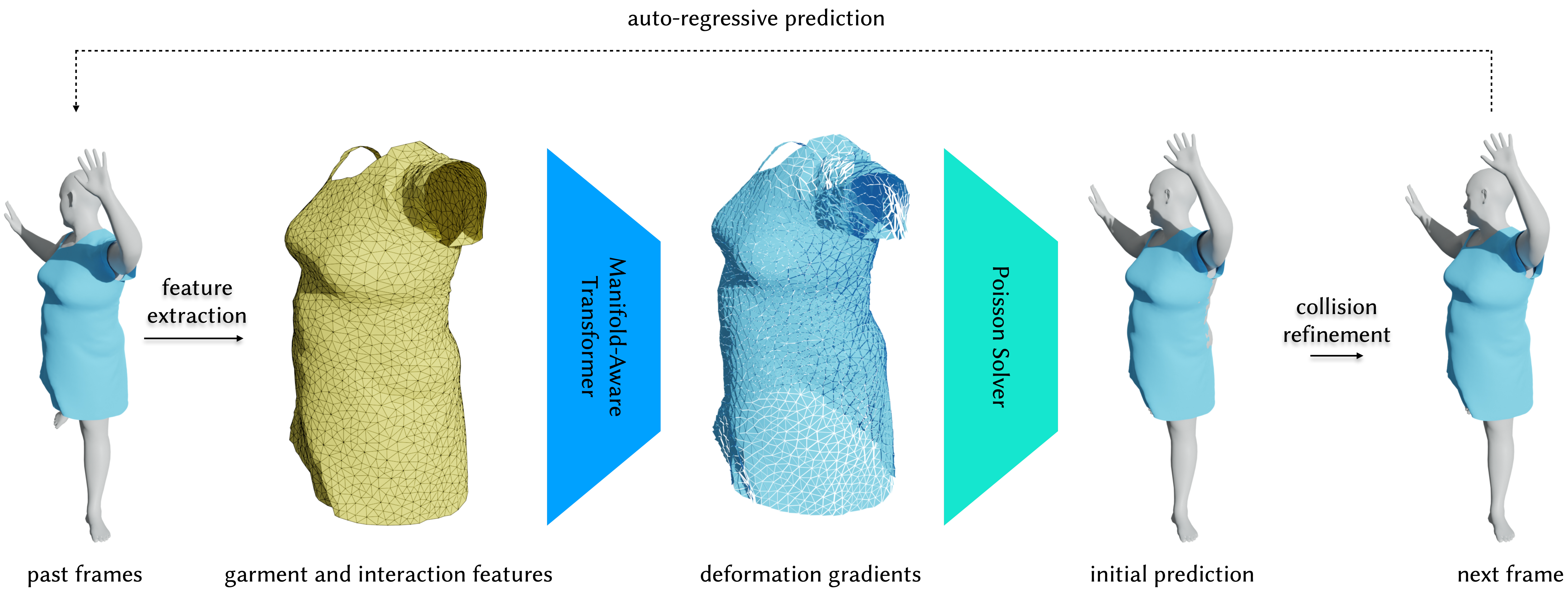}
    \caption{
    Our framework overview. We extract the garment features and interaction features on the garment geometry from the past frames. Our manifold-aware transformer is then applied spatially to the input features and predicts the relative deformation gradients to the next frame. An initial prediction is obtained with a Poisson solver. After the collision refinement, we get the prediction for the next frame. We auto-regressively repeat this process until the desired number of frames is reached.}
    \label{fig:overview}
  \end{figure*}

\subsection{Data-driven and Learning-based Methods}

Prior to the bloom of deep learning, there have been several data-driven approaches to model garment deformations. Aguiar et al.~\cite{de2010stable} propose to learn a linear dynamic system on the PCA subspace of garment deformation driven by a pre-defined body to achieve real-time performance. Guan et al.~\cite{guan2012drape} also explore statistical models to tackle how garments drape on different body shapes and poses. Luo et al.~\cite{luo2018nnwarp} use a lightweight neural network to transform a linear elasticity-based deformation into a non-linear deformation. Holden et al.~\cite{holden2019subspace} introduce neural networks that adapt subspace representations, making it possible to handle interactions between multiple objects. While being efficient, subspace-based methods are often limited to the training data and difficult to extrapolate to unseen settings. 

In order to leverage the recent success of neural network architectures in the 2D domain, recent approaches have utilized 2D canonical representations (i.e., UV mapping) for encoding the garment deformation.
DeepWrinkle~\cite{lahner2018deepwrinkles} predicts pose-dependent wrinkles represented as normal maps in the UV space. Zhang et al.~\cite{zhang2021deep} propose to refine the details of coarse simulation using a similar representation. In addition to pose-dependent effects, Jin et al.\cite{jin2020pixel} aim to model motion-dependent deformations.
While such UV-based representations effectively utilize 2D convolutions, they are limited in encoding spatial neighborhoods across UV seam boundaries.

Predicting the 3D geometry of the garment directly is considered an alternative solution. Gundogdu et al.~\cite{gundogdu2019garnet}, Habermann et al.~\cite{habermann2021real} fuse body and garment geometry features to predict pose-dependent effects in the canonical pose, and deform the result with linear blend skinning (LBS). Patel et al.~\cite{patel2020tailornet} also incorporate the dynamics and the change of garment styles, but model the garment as a height field over the body surface which is limited to tight-fitting garments.
Zhang et al.~\cite{zhang2022motion} handle loose garments by first learning a plausible deformation latent space but still training garment-specific networks. Pan et al.~\cite{pan2022predicting} utilize a virtual skeleton with additional virtual joints to better capture the low-frequency of loose garments with respect to the body motion. Similar to D. Li et al.~\cite{d2022n}, high-frequency displacements are then added with a graph neural network. While showing impressive progress, many of these methods, however, are specific to a training garment.

The time-consuming generation of training data with physically-based simulation often acts as a bottleneck for data-driven approaches. Hence, several unsupervised methods have been proposed recently. 
Bertiche et al.~\cite{bertiche2021pbns} use physically-inspired loss terms in combination with LBS deformation to predict pose-dependent deformations. A similar approach is also used by De Luigi et al.~\cite{de2022drapenet}. Santesteban et al~\cite{santesteban2022snug} extend the idea to and introduce strain and inertia-based losses. 
\cite{bertiche2022neural} train a network to predict the garment status such that it minimizes a combination of energy terms. We provide comparisons to some of these methods in the experiments section.

While learning-based methods have shown remarkable advances in recent years, generalization, i.e., generalizing to unseen garment types, still remains a challenge. In most recent concurrent approaches, GarSim~\cite{tiwari2023garsim} and HOOD~\cite{grigorev2023hood} tackle this challenge by utilizing a graph-convolution-based framework~\cite{pfaff2020learning}. Due to the limited receptive field of graph convolutions, their methods are sensitive to the resolution of the input meshes and require specific designs to incorporate more global information. We provide qualitative comparisons in the supplementary material regarding these methods. 

\subsection{Transformers}

Proposed by Vaswani et al.~\cite{vaswani2017attention} for natural language processing, transformers have quickly been adapted to many areas, including 3D tasks~\cite{guo2021pct,dwivedi2020generalization,chandran2022shape, ying2023adaptive}. We also adapt a transformer-based architecture and introduce a \emph{manifold-aware} structure to capture both local mesh neighborhoods and long-distance dependencies effectively.
\section{Overview}

Given the motion of an underlying collider, such as a human body, our goal is to predict the deformation of the garment as it interacts with the body. Our key insight is that garment deformations can be predicted in a generalizable manner by modeling the local interaction between the garment and the body surface. We introduce a set of garment features (e.g., the deformation gradient, velocity, and relative distance between the garment and the body) that capture such local interactions. We form tokens from the features obtained from a set of triangle faces sampled on the garment surface mesh. We introduce a manifold-aware transformer network that utilizes such tokens obtained from a set of past frames to predict the deformation of the garment in the current frame. We empower the transformer network to be \emph{manifold-aware} by encoding the geodesic information of the garment in addition to the local interaction features. Specifically, we replace part of the learned attention weights with the geodesic matrix obtained from the garment surface.

We represent the garment deformation using a Jacobian field~\cite{aigerman2022neural}. We discretize the continuous Jacobian field with random samples obtained on the mesh representation of the garment. Given the predicted deformation gradient, we solve a Poisson equation to reconstruct the explicit garment surface. Our approach is agnostic to the connectivity and the density of the triangulation of the garment surface and can handle garments with various topologies. We provide the overall architecture of our framework in \Cref{fig:overview} and next discuss the details of our approach.

\section{Method}

\begin{figure*}[t!]
    \newcommand{\infe}{\color[rgb]{0.99,0.72,0.18}}
    \newcommand{\embedding}{\color[rgb]{0.97,0.74,0.84}}
    \newcommand{\outfe}{\color[rgb]{0.07,0.69,0.15}}
    \centering
    \includegraphics[width=\linewidth]{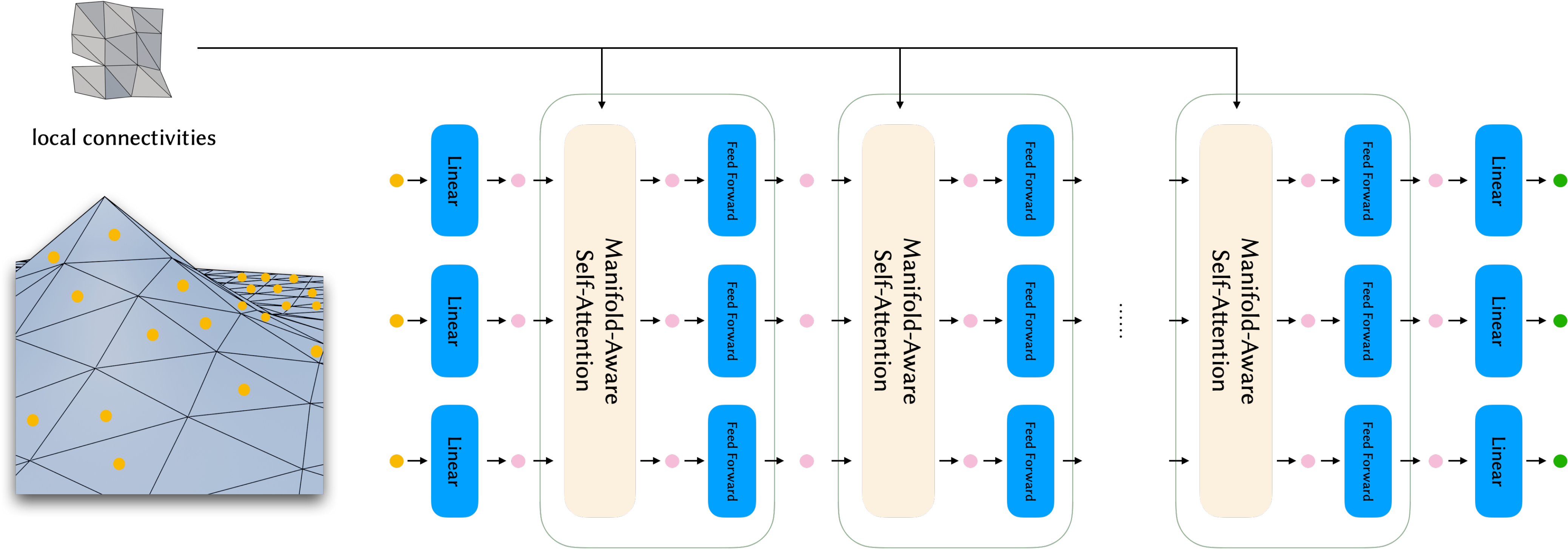}
    \caption{
    Our manifold-aware transformer architecture. The {\infe {input features}} are extracted from the faces of the input mesh. After being projected into {\embedding {embedding space}} by a linear transformation, they are fed into the transformer encoder consisting of $n_l$ identical layers. Our manifold-aware self-attention layers explicitly involve local connectivities of the input geometry, making it possible to predict accurate dynamics caused by seams. The output of the transformer encoder is projected to the {\outfe {output features}} by another linear transformation. The output features are then used to predict the next frame of garment deformation.}
    \label{fig:architecture}
  \end{figure*}
\subsection{Pipeline}
\label{sec:inference}

We represent the garment and the underlying body geometry with their corresponding vertex positions at a particular time $t$ and the mesh triangulation. Specifically, let $\{\bbv^{t},\bbt^{g}\}$ denote the garment with triangulation $\bbt^{g}$ and vertex positions $\bbv^{t}$ and let $\{\bbu^{t},\bbt^{b}\}$ denote the underlying body with triangulation $\bbt^{b}$ and vertex positions $\bbu^{t}$. Given the previous states of the garment in the past $\nhist$ frames, i.e., $\{\bbv^{t-\nhist+1},\cdots,\bbv^{t}\}$, and the state of the body in the next frame, i.e., $\bbu^{t+1}$, our goal is to predict $\bbv^{t+1}$, i.e., the state of the garment in the next frame. We introduce a transformer-based network that takes as input a set of features $\bbF^t_i$ computed for each face on the garment surface utilizing the past and current states of the garment and the underlying body. 
The output is the relative deformation gradients $\Psi^{t+1}$ of each face in the next frame, and global velocity $\bbq^{t+1}$ of the entire geometry. We then compute the absolute deformation $\Phi^{t+1}$ to reconstruct the garment mesh at time $t+1$ via the Possion equation~\cite{sumner2004deformation, sorkine2004laplacian}. To reduce the accumulation of error, we also predict the singular values $\Sigma^{t+1}$ for $\Phi^{t+1}$ at the next frame and use it to regularize the stretching. Since the predicted cloth geometry is not guaranteed to be collision-free with the body geometry, we use a post-processing strategy adopted from DRAPE~\cite{guan2012drape}. A pseudo-code of this process is shown in \Cref{alg:predict_next}.
We refer the readers to the supplementary material for a detailed discussion on the deformation gradients and the post-process collision refinement method.

\begin{algorithm}
    \caption{Prediction of frame $t+1$ from the past $\nhist$ frames}
    \label{alg:predict_next}
    \begin{algorithmic}
    \State \textbf{Input:} Garment vertex positions $\bbv^{t-\nhist:t}$, body vertex position $\bbu^{t-\nhist:t+1}$
    \State \textbf{Output:} Vertex positions $\bbv^{t+1}$
    
    \Procedure{PredictFrame}{$\bbv^{t-\nhist:t}, \bbu^{t-\nhist:t+1}$}
        \State $\bbF^t \gets$ features extracted from garment $\bbv$ and body $\bbu$
        \State $\Psi^{t+1}$, $\Sigma^{t+1}$, $\bbq^{t+1} \gets$ prediction from the network with $\bbF^t$
        \State $\bar\Phi^{t+1} \gets \Psi^{t+1}\Phi^t$
        \State $\Phi^{t+1} \gets$ replace singular values of $\bar\Phi^{t+1}$ with $\Sigma^{t+1}$
        \State $\bar\bbv^{t+1} \gets$ solve a Poisson problem with $\Phi^{t+1}$ and velocity $\bbq^{t+1}$
        \State $\bbv^{t+1} \gets$ collision refinement for $\bar\bbv^{t+1}$
    \EndProcedure
    \end{algorithmic}
\end{algorithm}

\subsection{Garment and Interaction Features}
Our network takes as input a set of features defined on the garment geometry. In order to capture the dynamics of the garment, we stack the features obtained from past $\nhist$ frames together. For simplicity, we describe the features computed from a single frame and omit the frame index $t$ in the following text unless otherwise specified. 

In addition to the deformation gradient $\Phi_i$, derived from the predictions of preceding frames, in an auto-regressive fashion into the present frame, we also define a set of features for a given triangle $i$ to encode the current state of the garment as well as its interaction with the body. The following features encode the state of the garment geometry.

\textbf{Orientation.} The deformation gradient $\Phi_i$ records the  deformation \emph{relative} to the rest state. Hence, the network is not aware of the orientation of the surface. To mitigate this issue, we include the normal direction $\bbn_i$ in the world-coordinate of triangle $i$ as part of the input feature. 

\textbf{Centroid.} The input to the transformer network is permutation invariant. Similar to the positional encoding used in the original transformer, we include the centralized centroid coordinate $\bbc_i = c_i - z$ of each triangle to preserve the spatial order information. $c_i$ is the centroid of triangle $i$ and $z = 1/|T| \sum_{i \in T} c_i$ is the average centroid of the garment at a given frame.

\textbf{Global velocity.} The solution of the Poisson equation is not unique up to a translation. We thus incorporate the per-frame global velocity $\bbq^t = z^t - z^{t-1}$ as part of the input.

In order to capture the interaction of the garment with the underlying body, we also define a set of \emph{interaction} features as follows:

\textbf{Signed distance.} We encode the relative position of the garment with respect to the body. For every triangle $i$ in the garment, we record its signed distance $d_i$ to the body. In addition, we also record a direction $\vec v_i$ of the nearest point on the body to the centroid $c_i$ of triangle $i$. This constitutes the singed distance feature $\bbs_i = \{d_i, \vec v_i\} \in \mathds{R}^4$. It serves as the local coordinate of the collider, enabling us to encode collider deformation in the garment space.

\textbf{Collider deformation.} For the network to predict the deformation at frame $t + 1$, we incorporate the deformation of the collider, i.e., the body, at frame $t + 1$ as part of the input feature. To this end, for each face $i$ in the garment in the current frame, we compute its nearest face on the body in the current frame. We define a \emph{collider deformation feature}, $\bbd_i = \{d_i, \vec q_i\} \in \mathds{R}^{12}$, where $d_i \in \mathds{R}^{3 \times 3}$ represents the relative deformation gradient to the next frame of its nearest body face and $\vec q_i$ is the velocity of the centroid of the nearest body face.

Note that, unlike commonly used SMPL~\cite{SMPL2015} body and pose parameters, the proposed interaction features are not limited to a specific body model and can be directly applied to any other type of collider geometry. We conduct experiments showing the versatility of the proposed interaction features in \Cref{sec:collider}.

\subsection{Manifold-aware transformer networks}

\textbf{Network input and output.} We denote the concatenation of the aforementioned per-triangle features as $\bbf_i = \{ \Phi_i, \bbn_i, \bbc_i, \bbs_i, \bbd_i, \bbsigma_i\}$. Furthermore, we collect the features of past $\nhist$ frames together as $\bbF^t_i = \{\bbf_i^k \}_{k=t - \nhist+1}^t$. The input to our network is $\bbF^t$ along with the global velocity of the garment in the past $\nhist-1$ frames denoted as $\bbQ^t = \{\bbq^k\}_{k=t - \nhist + 2}^t$. The network then predicts the relative deformation gradient $\Psi_i^{t+1}$ and the singular value $\bbsigma_i^{t+1}$ for every face $i$ and the global velocity $\bbq^{t+1}$ of the garment in the next frame.

\textbf{Architecture.} The overview of our network architecture is demonstrated in \Cref{fig:architecture}. For memory and computational efficiency, we evenly split the faces into $n_s$ disjoint subsets $\{\bbt_i\}_{i=1}^{n_s}$. 
For each of the split $\bbt_i \subset \bbt$, we gather the features $\{\bbF_j^t\}_{j \in \bbt_i}$ of every triangle in this split, concatenated with the global velocity feature $\bbQ^t$, as the input to the network. Note that the features $\{\bbF_j^t\}$ are calculated before the splitting, and no downsampling is involved for feature calculation. A linear transformation first maps the input features into an embedding space of dimension $n_e$. The embeddings of the features are then passed through $n_l$ transformer~\cite{vaswani2017attention} encoder layers, where the self-attention mechanism learns to capture the global context. Unlike graph convolution-based networks with only limited receptive fields, our transformer-based architecture is capable of learning long distance correlation. The output of the last encoder layer is passed through a linear layer to predict the relative deformation gradient $\Psi^{t+1}$ and singular values $\bbsigma^{t+1}$, as well as the global velocity $\bbq^{t+1}$ for the next frame. 
Note that this framework is triangulation-agnostic and is by nature not limited to a single garment. The network can be trained with multiple types of garments and colliders, and can be used to predict unseen clothes. We demonstrate the versatility of our network in \Cref{sec:main-result}.

\begin{figure}
    \centering
    \newcommand{\pll}{-5}
    \begin{overpic}[width=\columnwidth]{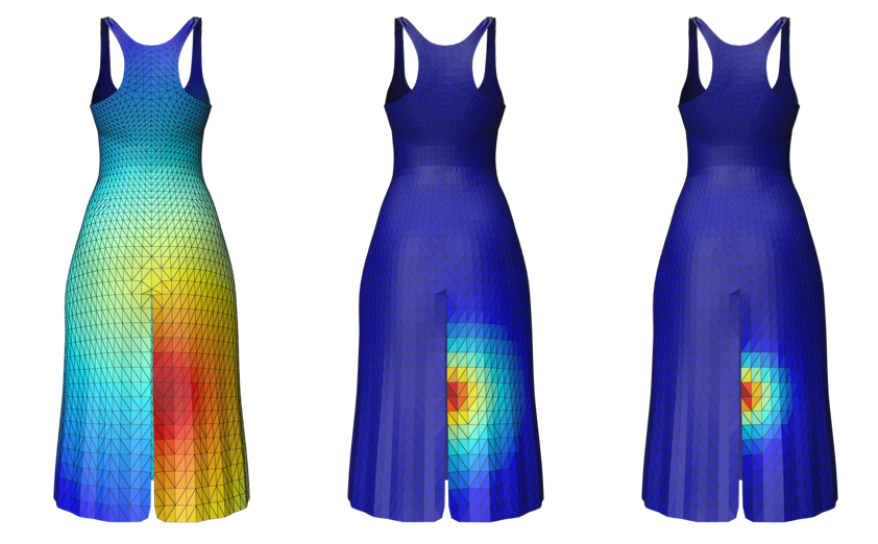}
      \put(11,  \pll){{\small $\pgeo = 1$}}
      \put(45,  \pll){{\small $\pgeo = 5$}}
      \put(78,  \pll){{\small $\pgeo = 20$}}
      \end{overpic}
    \caption{Geodesic attention weights. We use $\pgeo$ to control the attention range.}
    \label{fig:geodesic}
  \end{figure}

\textbf{Manifold-aware self-attention.} For the cases with complex garments or challenging body poses, close-by spatial locations over the garment surface may have very different dynamic behavior, e.g., the sleeve opening for left/right arm of a shirt can be spatially close under a crossed arm pose while their dynamic behavior might be different as shown in \Cref{fig:manifold-aware}. To prevent such spurious correlations, we consider the connectivity of the garment surface. Specifically, we propose a manifold-aware self-attention mechanism. For $\nconn$ heads, we use the pairwise geodesic distance between the centroids of the triangles as the attention score as below:
\begin{equation}
    \bbA_{i\cdot} = \mathrm{softmax}(-\bbD_{i\cdot}^{\pgeo}),
    \label{eq:connectivity-attention}
\end{equation}
where $\bbD_{ij}$ denote the matrix of pair-wise geodesic, $\pgeo$ is the exponential index to control the attention range. We visualize the geodesic attention weights for a triangle in \Cref{fig:geodesic}. For an in-depth study of the effect of manifold-aware self-attention, we refer the readers to \Cref{sec:manifold-aware}.

\subsubsection{Singular value prediction}

Our network predicts the relative deformation between two consecutive frames and auto-regressively predicts further frames based on existing predictions. 
This can lead to error accumulation of the absolute deformation gradient, and cause severe area distortion of the geometry. As singular values indicate the scaling of the deformation gradient along the three principal directions, to mitigate this issue, we also predict the singular values of the deformation gradient.
During inference time, we accumulate the predicted relative deformation gradients to absolute deformation gradients. We perform SVD and replace the singular values with the predicted ones before using them to reconstruct the deformed mesh with Poisson equation. We refer the readers to \Cref{sec:results} for the study on the effect of singular value prediction.

\subsubsection{Loss functions and training} 

We adopt a fully-supervised training scheme to learn the deformation field from a set of physically-simulated training data. The training is supervised via the following loss terms:

\textbf{Deformation gradient loss.} 
The L1 norm is employed to measure the difference between the predicted relative deformation gradients $\Psi_i^{t+1}$ and the ground truth $\tilde\Psi_i^{t+1}$:
\begin{equation}
    \Loss_{\text{def}} = \frac{1}{|\bbt_i|}\sum_{j \in \bbt_i} \| \Psi_j^{t+1} - \tilde\Psi_j^{t+1} \|_1.
    \label{eq:loss_def}
\end{equation}

\textbf{Singular value loss.} 
Similarly, the L1 norm is utilized to measure the difference between the predicted singular values  $\bbsigma_i^{t+1}$ and the ground truth $\tilde\bbsigma_i^{t+1}$:
\begin{equation}
    \Loss_{\text{sv}} = \frac{1}{|\bbt_i|}\sum_{j \in \bbt_i} \| \bbsigma_j^{t+1} - \tilde\bbsigma_j^{t+1} \|_1.
    \label{eq:loss_sv}
\end{equation}

\textbf{Global velocity loss.} 
The L1 norm is applied to measure the difference between the predicted global velocity $\bbq^t$ and the ground truth $\tilde\bbq^t$:
\begin{equation}
    \Loss_{\text{vel}} = \| \bbq^t - \tilde\bbq^t \|_1.
    \label{eq:loss_vel}
\end{equation}
Our full loss used for training summarizes as:
\begin{equation}
    \Loss = \Loss_{\text{def}} + \lambda_{\text{sv}}\Loss_{\text{sv}} + \lambda_{\text{vel}}\Loss_{\text{vel}}.
    \label{eq:loss_overall}
\end{equation}
During training, we supervise only one step of prediction. We random sample $\nhist + 1$ consecutive frames from a physically-simulated dataset and use the features of the first $\nhist$ frames as input of the network, and supervise the prediction on frame $\nhist + 1$ using the loss functions in \Cref{eq:loss_overall}. Besides, all input features are normalized to have zero mean and unit variance. To prevent error accumulation in our auto-regressive workflow, we add noise to the input features of the network during the training. Specifically, we add a random noise $\varepsilon \sim \mathcal{N}(0, \sigma_n)$ to the normalized input features $\bbF^t$. The noise is sampled independently for each feature dimension and each triangle. We find that this is a simple yet effective solution to stabilize the long-term generation. 
For a detailed description of the layers in our network and the specific values of the hyper-parameters, we refer to the supplementary material.

\section{Experiments}
\label{sec:experiments}

\begin{figure}
    \centering
    \includegraphics[width=\linewidth]{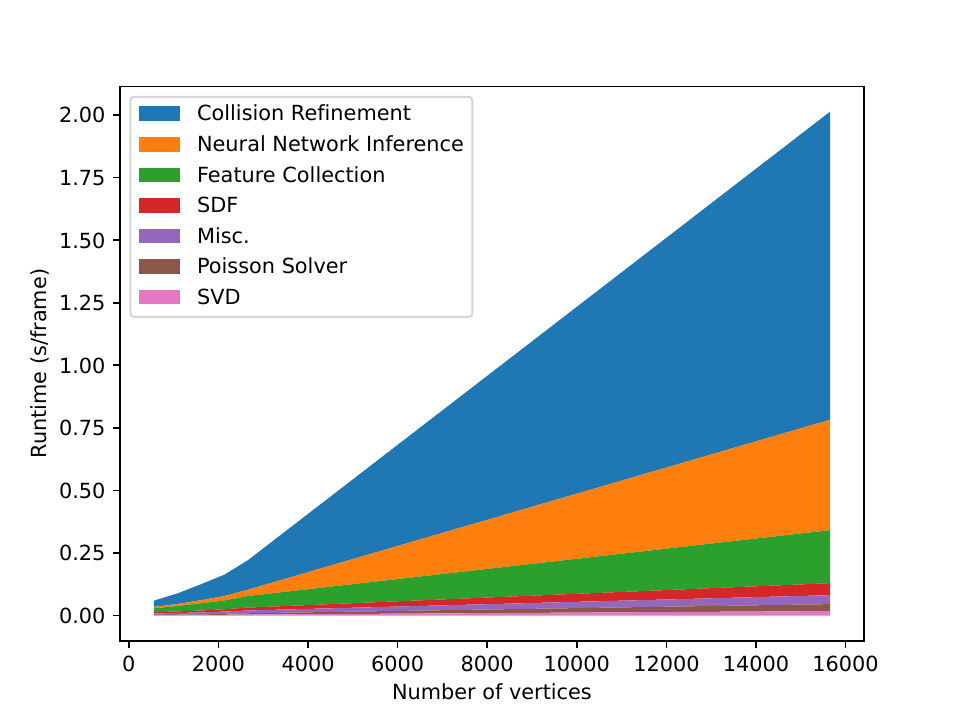}
    \caption{Breakdown of inference performance.}
    \label{fig:inference-time}
  \end{figure}

We evaluate our approach on various garment and body types to demonstrate the generalization ability of our method. We also compare with other neural techniques and provide an ablation to evaluate the effectiveness of various components in our design. Please refer to the supplementary video for qualitative results.

\subsection{Implementation details}
Our framework is implemented in PyTorch~\cite{NEURIPS2019_9015}, and the experiments are performed on an NVIDIA GeForce RTX 3090 GPU. We optimize the parameters of our network with the loss term in \Cref{eq:loss_overall}
using the Adam optimizer~\cite{kingma2014adam}. It takes about 48 hours to train our network. We refer the readers for a detailed description of our network architecture and hyper-parameters to the supplementary material.

\textbf{Running time.} We show a breakdown of the running time of each component in \Cref{fig:inference-time}. The most expensive operation is collision refinement, which involves solving a sparse linear system of size $3N \times 3N$, where $N$ is the number of vertices. The feature collection contains the computation of the network input $\bbf_i$ excluding signed distance function (SDF), which is listed separately. Our efficient SDF calculation is implemented on GPU by combining minimum pairwise distance and winding numbers~\cite{jacobson2013robust}. The Poisson solver is implemented with a pre-computed Cholesky decomposition using CHOLMOD~\cite{cholmod} and solving on the GPU~\cite{naumov2011parallel} using the implementation by Nicolet et al.~\cite{nicolet2021large}.

\begin{figure}
    \centering
    \newcommand{\pll}{-5}
    \begin{overpic}[width=\linewidth]{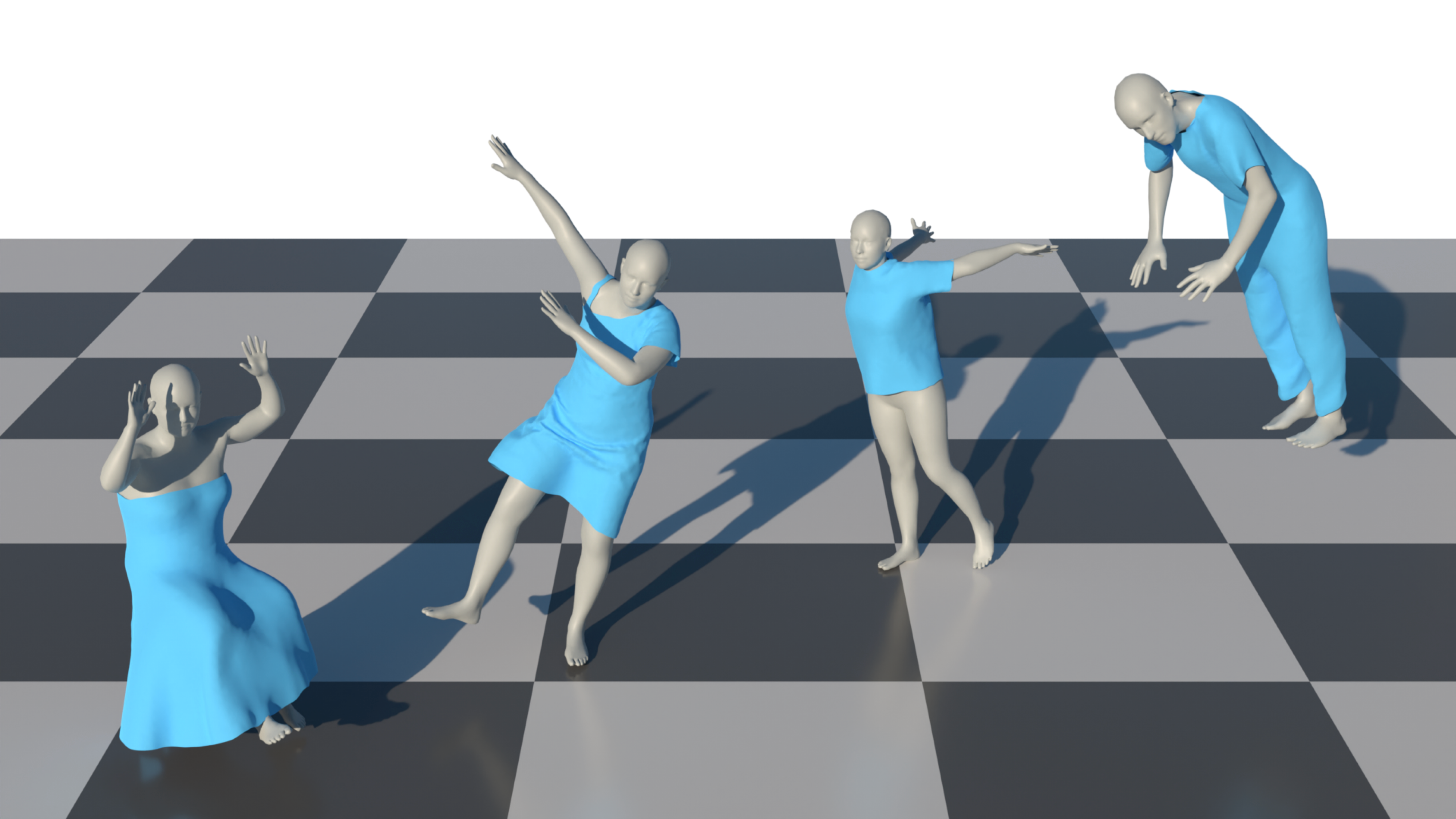}
      \end{overpic}
    \caption{A general model for different garments. Our model is capable of predicting the dynamics of different garments driven by various motions.}
    \label{fig:different-garments}
  \end{figure}

\textbf{Dataset.}
\label{sec:main-result}
We train our model with the CLOTH3D dataset~\cite{bertiche2020cloth3d} which contains 7 categories of garments (shirt, shirt, top, trousers, skirt, jumpsuit, and dress). Within each category, garment shape is augmented using cutting and resizing. The garments are then simulated in 3D on a body that is animated with a motion sequence from the CMU human motion dataset~\cite{CMUmocap}. We randomly select 200 simulated sequences (around 50,000 frames) and train our model with 6 garment categories, excluding the ``skirt'' category. Since the dataset also contains several material configurations, we use the ``cotton'' configuration as our training data. 

At test time, we evaluate our model with unseen augmentations (i.e., cutting and resizing) of the seen garment types and with garments in the unseen ``skirt'' category, driven by 30 unseen motion sequences.

\textbf{Evaluation Metrics.} We evaluate our method with respect to ground truth simulation results using the mean vertex error (in cm) and Chamfer Distance~\cite{wu2021density}. We also measure the geodesic distortion by calculating the L1 error between the pair-wise geodesic distance of the generated results and ground truth.

\subsection{Results}
\label{sec:results}
As shown in \Cref{fig:different-garments} and our supplementary video, our method can synthesize realistic results on different types of garments and correctly capture the subtle dynamics. In the following, we provide a more thorough evaluation.

\textbf{Robustness to remeshing.}
\begin{figure}
    \centering
    \newcommand{\pll}{-5}
    \begin{overpic}[width=\columnwidth]{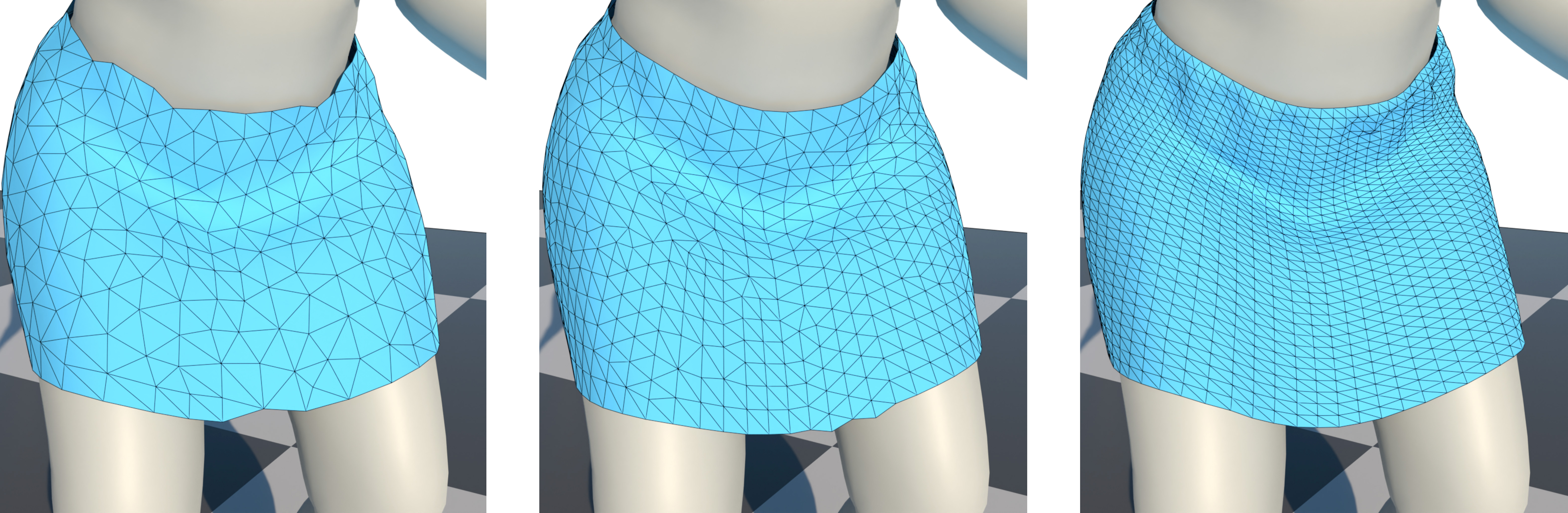}
      \put(10,  \pll){{\small 1032 faces}}
      \put(44,  \pll){{\small 2067 faces}}
      \put(78,  \pll){{\small 5168 faces}}
      \end{overpic}
    \caption{Robustness to remeshing. Our network predicts consistent results for different meshing, thanks to our feature representation and network architecture.}
    \label{fig:resolution}
  \end{figure}
The input and output features of our network are triangulation-agnostic. Furthermore, the manifold-aware self-attention module is also robust to changes in triangulation. We evaluate our model on the same garment with different meshing, specifically containing 1032, 2067, and 5168 faces, respectively. It can be seen in \Cref{fig:resolution} that our method generates consistent results across different mesh resolutions, which is not possible to handle with graph-convolution-based methods~\cite{grigorev2023hood, tiwari2023garsim}.

\begin{figure}
    \centering
    \newcommand{\pll}{-5}
    \begin{overpic}[width=\columnwidth]{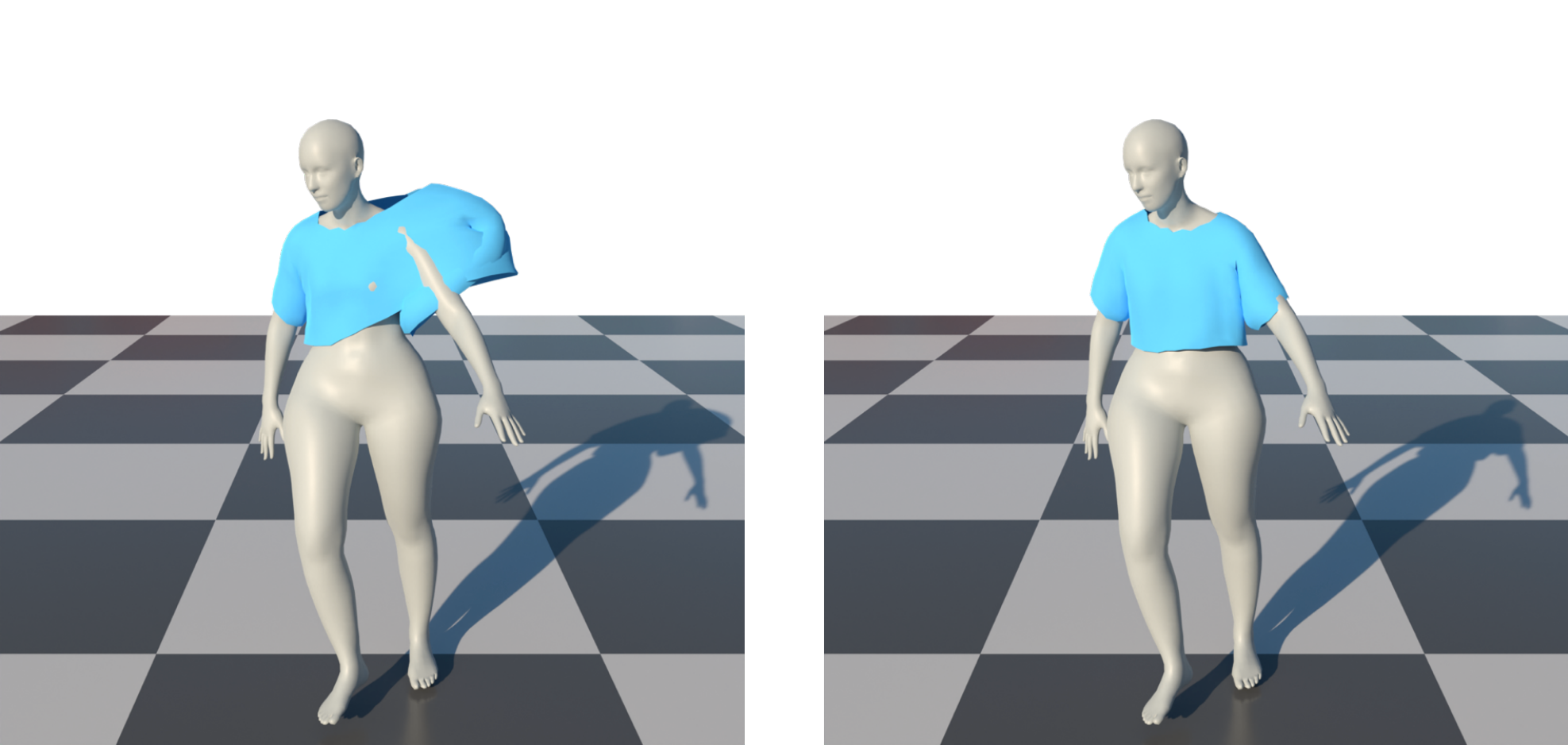}
      \put(9,  \pll){{\small w/o manifold awareness}}
      \put(68,  \pll){{\small Full approach}}
      \end{overpic}
    \caption{Manifold awareness. The manifold aware component is able to resolve instances where spatial proximity happens to unconnected cloth segments. While without the manifold-aware component, it fails to distinguish the two segments.}
    \label{fig:manifold-aware}
  \end{figure}
\textbf{Effect of manifold-aware self-attention.}
\label{sec:manifold-aware}
In certain motion sequences, two different parts of the garment (e.g., the sleeves and the torso) can spatially come close together while their dynamics are still significantly different. As shown in \Cref{fig:manifold-aware}, our manifold-aware attention module can effectively handle such cases and generate plausible results.

\begin{figure}
    \centering
    \newcommand{\pll}{-5}
    \begin{overpic}[width=\columnwidth]{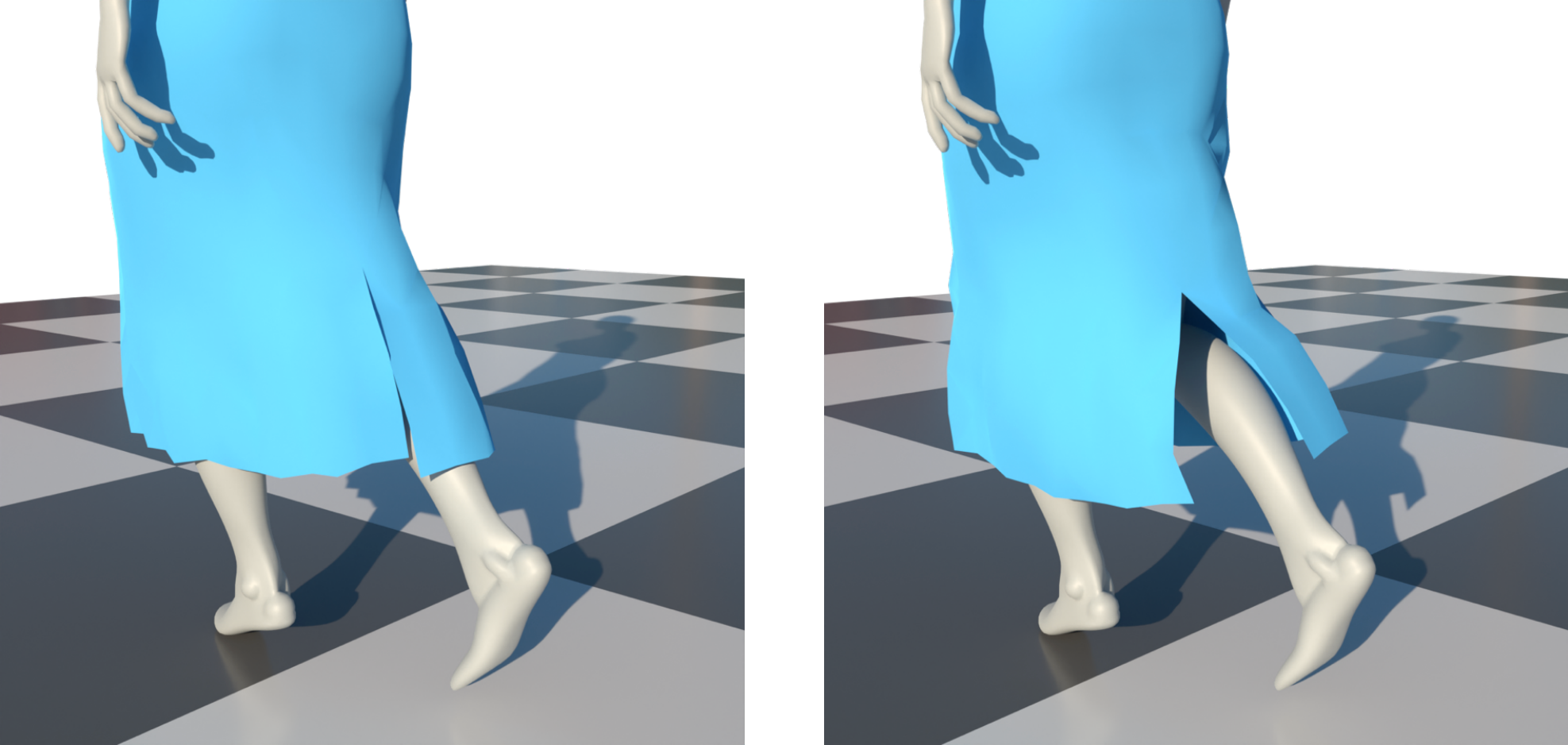}
      \put(8,  \pll){{\small w/o Manifold wareness}}
      \put(67,  \pll){{\small Full approach}}
      \end{overpic}
    \caption{Local connectivity changes, such as a cut as seen here, are incorporated into our network's prediction with the help of the manifold-aware transformer.}
    \label{fig:dress-cut}
  \end{figure}

Since our model explicitly encodes the geodesic information, it can generalize to garments with unseen seam cuts. Due to the lack of connectivity changes in the CLOTH3D dataset, we created a dataset from a given dress with 10 different seam cuts as the training set and 2 different seam cuts as the test set. It can be seen in \Cref{fig:dress-cut} that our model is able to synthesize plausible results reflecting the unseen seam cuts without re-training. Please refer to the accompanying video for a complete result.

\begin{table}
    \caption{Ablation study.}
    \begin{tabular}{l c c}
    \toprule
    & {\small Mean vertex error (cm)} & {\small Geodesic distortion} \\
    \midrule
    {\small w/o manifold-aware} & 4.54 & $2.83 \times 10^{-2}$\\
    {\small w/o singular prediction} & {12.2} & {$9.33 \times 10^{-2}$} \\
    {\small Full approach} & 3.19 & $2.05 \times 10^{-2} $\\
    \bottomrule
    \end{tabular}
    \label{tab:qunat-ablation}
\end{table}
We conduct an ablation study over the manifold-aware component by using only learned attention weights. As shown in \Cref{tab:qunat-ablation}, the geodesic information not only leads to better visual performance, but also better preserves the geodesics.

\begin{figure}
    \centering
    \newcommand{\pll}{-5}
    \begin{overpic}[width=\columnwidth]{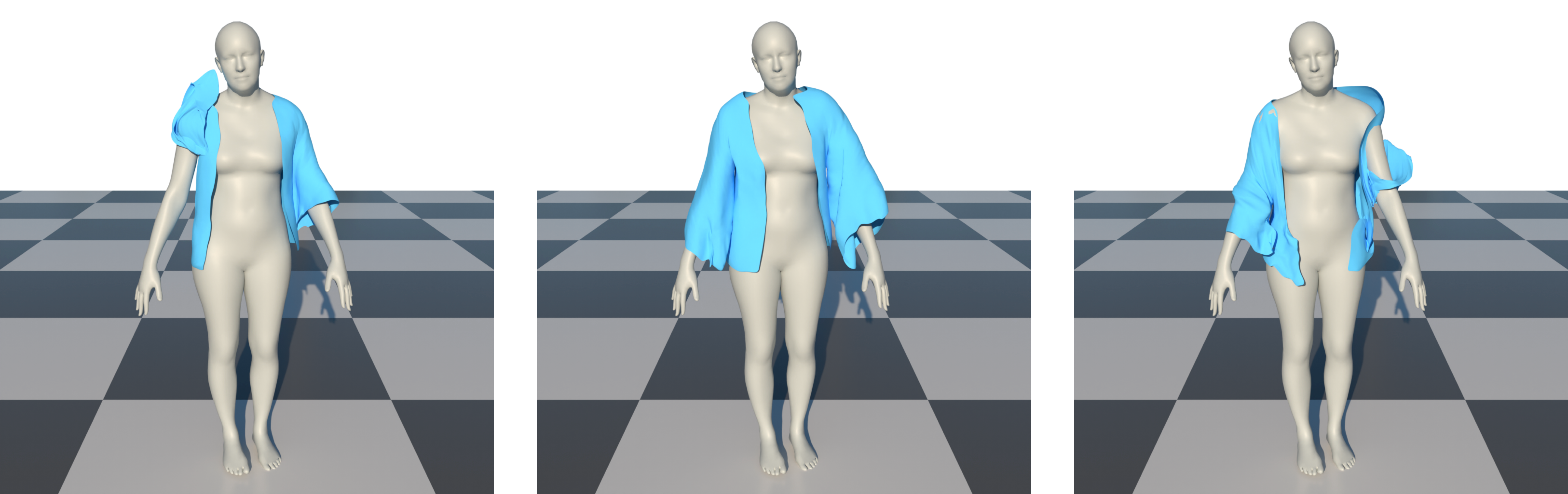}
      \put(10,  \pll){{\small Small $\pgeo$}}
      \put(43,  \pll){{\small Suitable $\pgeo$}}
      \put(78,  \pll){{\small Large $\pgeo$}}
      \end{overpic}
    \caption{Choice of $\pgeo$. We show that with a suitable choice of $\pgeo$, our method can handle the problem of spatial proximity, while a too small or a too large $\pgeo$ can lead to artifacts.}
    \label{fig:pgeo-cmp}
  \end{figure}
\textbf{Choice of $\pgeo$.} As can be seen in \Cref{fig:pgeo-cmp}, a too small or too large $\pgeo$ yields inferior results. We show that our model is robust to a large range of $\pgeo$ values in \Cref{tab:pgeo-ablation}. When $\pgeo$ ranges between 1 and 50, the mean vertex error (MVE) remains small. We use $\pgeo = 20$ in our other experiments.

\begin{table}
    \centering
    \caption{Ablation study on the choices of $\pgeo$.}
    \begin{tabular}{l c c c c c c c c}
    \toprule
    $\pgeo$ & 0.01 & 0.1 & 1 & 10 & 20 & 50 & 100 \\
    \midrule
    {\small MVE (cm)} & 4.34 & 3.79 & 3.23 & 3.31 & 3.19 & 3.20 & 4.72 \\
    \bottomrule
    \end{tabular}
    \label{tab:pgeo-ablation}
\end{table}

\begin{figure}
    \centering
    \newcommand{\pll}{-5}
    \begin{overpic}[width=\columnwidth]{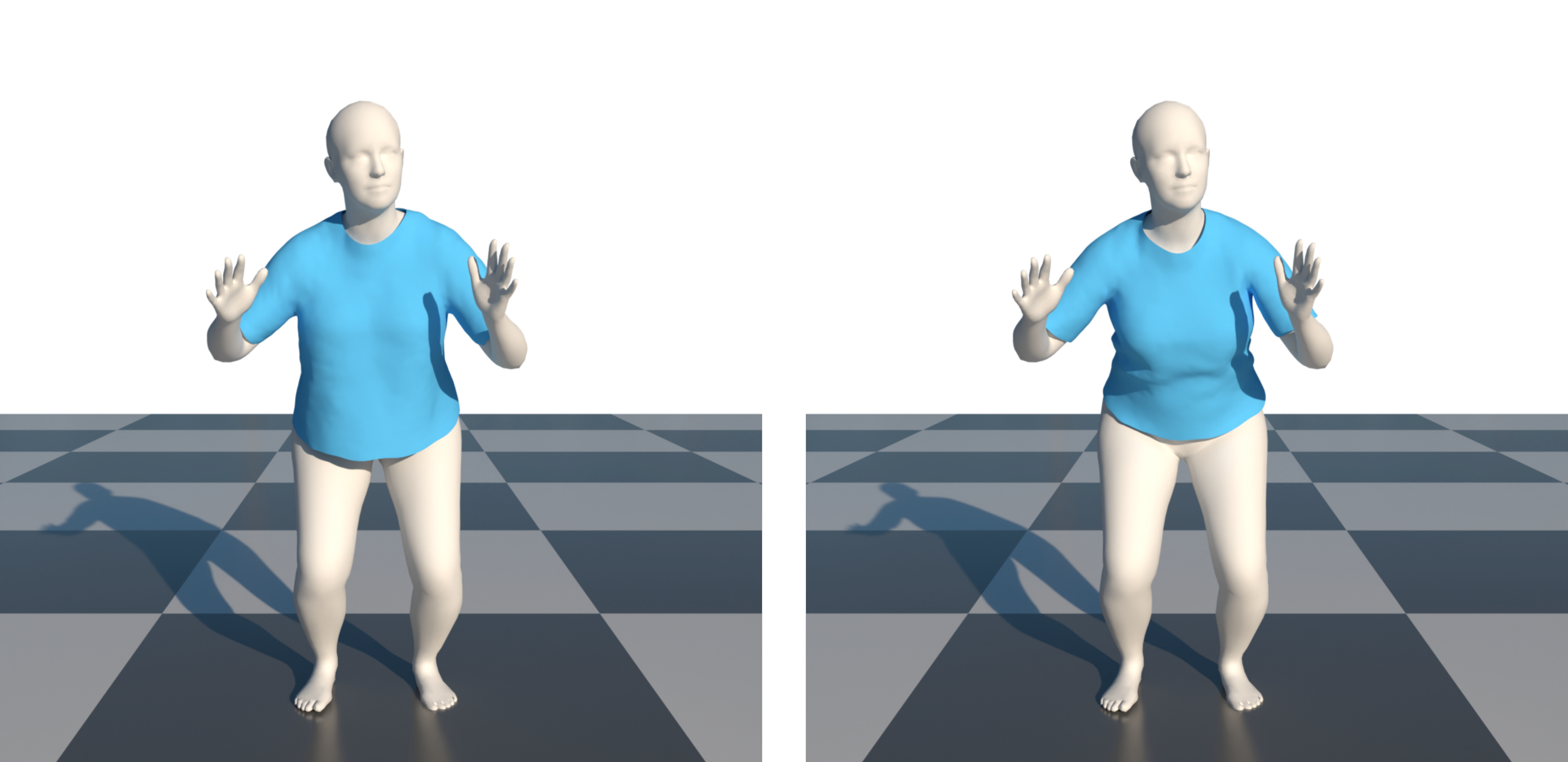}
      \put(5,  \pll){{\small w/o singular value prediction}}
      \put(68,  \pll){{\small Full approach}}
      \end{overpic}
    \caption{Singular value prediction. With the help of predicting singular values of absolute deformation gradient, our method does not suffer from the over-stretching problem.}
    \label{fig:sigma-cmp}
  \end{figure}

\textbf{Effect of singular value prediction.} We evaluate the effectiveness of our singular value prediction term in \Cref{eq:loss_sv} by removing it from the loss function and directly using the predicted deformation gradient without replacing the singular values. As shown in \Cref{tab:qunat-ablation} and \Cref{fig:sigma-cmp}, the singular value prediction term helps prevent the accumulation of stretching and produces more accurate results.

\begin{figure}
    \centering
    \newcommand{\pll}{-5}
    \begin{overpic}[width=\columnwidth]{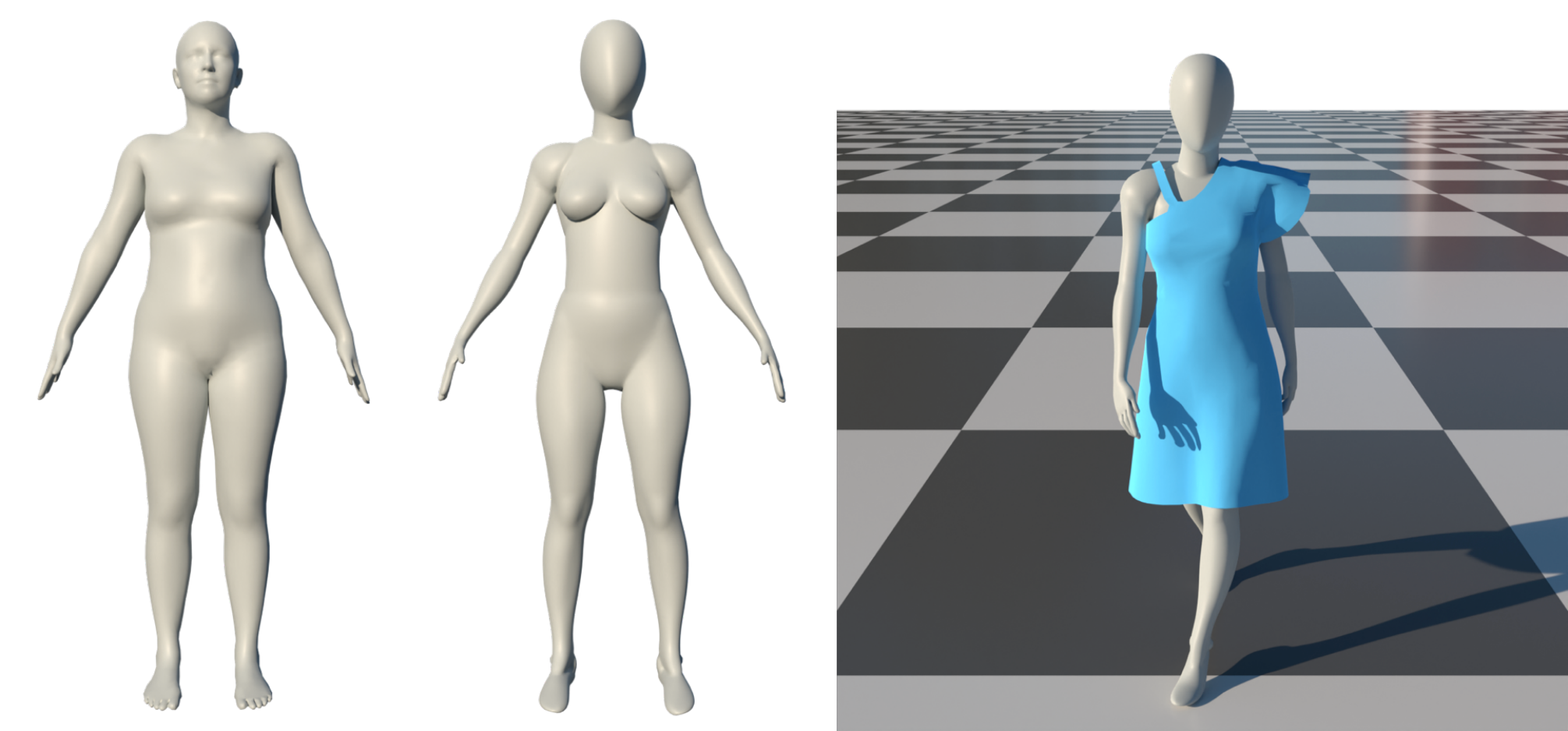}
      \put(4,  \pll){{\small Training body}}
      \put(31,  \pll){{\small Unseen body}}
      \put(70,  \pll){{\small Our result}}
      \end{overpic}
    \caption{Unseen body model. The mannequin wearing the dress is not seen by our model during training. However, it is still able to predict a plausible result. }
    \label{fig:unseen-body}
  \end{figure}

\textbf{Unseen body model.}
\label{sec:collider}
While our training data is driven by the SMPL~\cite{SMPL2015} model, our model can be applicable to different human body geometries. As shown in \Cref{fig:unseen-body}, we test our model with a mannequin, a significantly different body model from the SMPL and demonstrate plausible results. Please refer to the accompanying video for additional results.

\subsection{Comparisons}
\label{sec:comparisons}

\begin{table}
    \caption{Quantitative comparison to the supervised method.}
    \begin{tabular}{l c c}
    \toprule
    & {\small Mean vertex error (cm)} & {\small Chamfer distance} \\
    \midrule
    {\small SSCH~\cite{santesteban2021self}} & 2.93 & $3.84 \times 10^{-4}$ \\
    {\small Ours} & 2.69 & $3.30 \times 10^{-4}$ \\
    \bottomrule
    \end{tabular}
    \label{tab:qunat-supervised}
\end{table}
\begin{figure}
    \centering
    \newcommand{\pll}{-5}
    \begin{overpic}[width=\columnwidth]{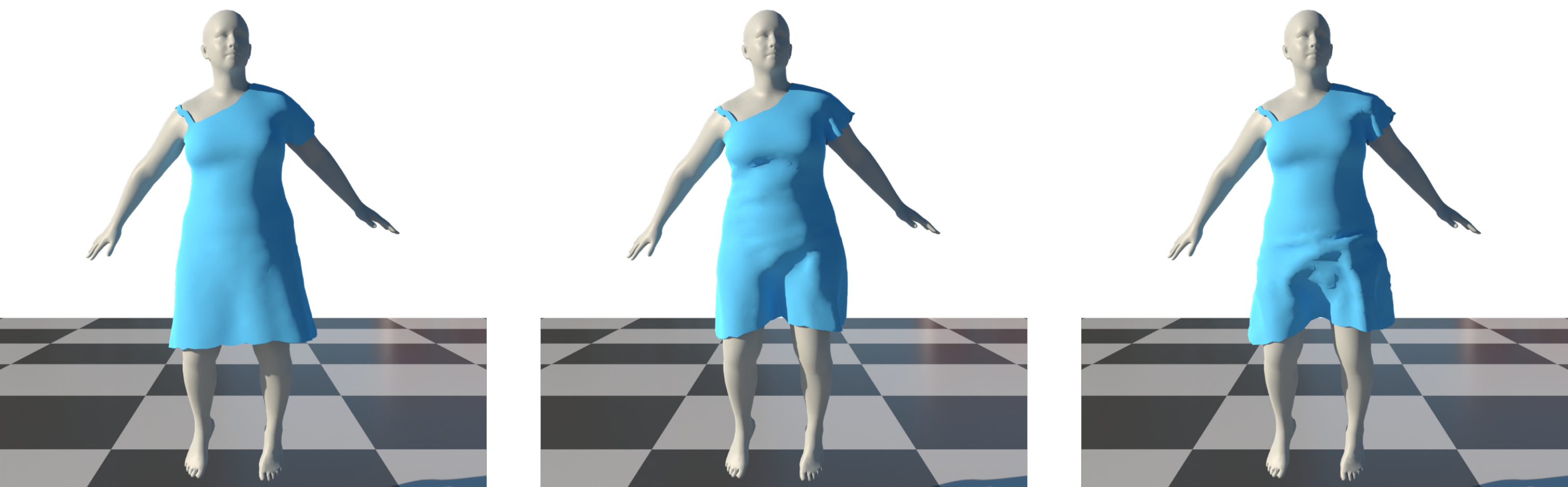}
      \put(5,  \pll){{\small SSCH~\cite{santesteban2021self}}}
      \put(46,  \pll){{\small Ours}}
      \put(75,  \pll){{\small Ground truth}}
      \end{overpic}
    \caption{Comparison to SSCH. Our network is able to faithfully capture the dynamics of the garment.}
    \label{fig:comp-supervised}
  \end{figure}
\textbf{Supervised methods.} We compare out method to SSCH~\cite{santesteban2021self} as a baseline supervised method. For a fair comparison, we retrain our model with the same VTO dataset this method is trained on. The results generated by our method faithfully reconstruct the dynamics of the garment as demonstrated in \Cref{fig:comp-supervised}. We also report superior quantitative results in \Cref{tab:qunat-supervised}.

\begin{figure}
    \centering
    \newcommand{\pll}{-5}
    \begin{overpic}[width=\columnwidth]{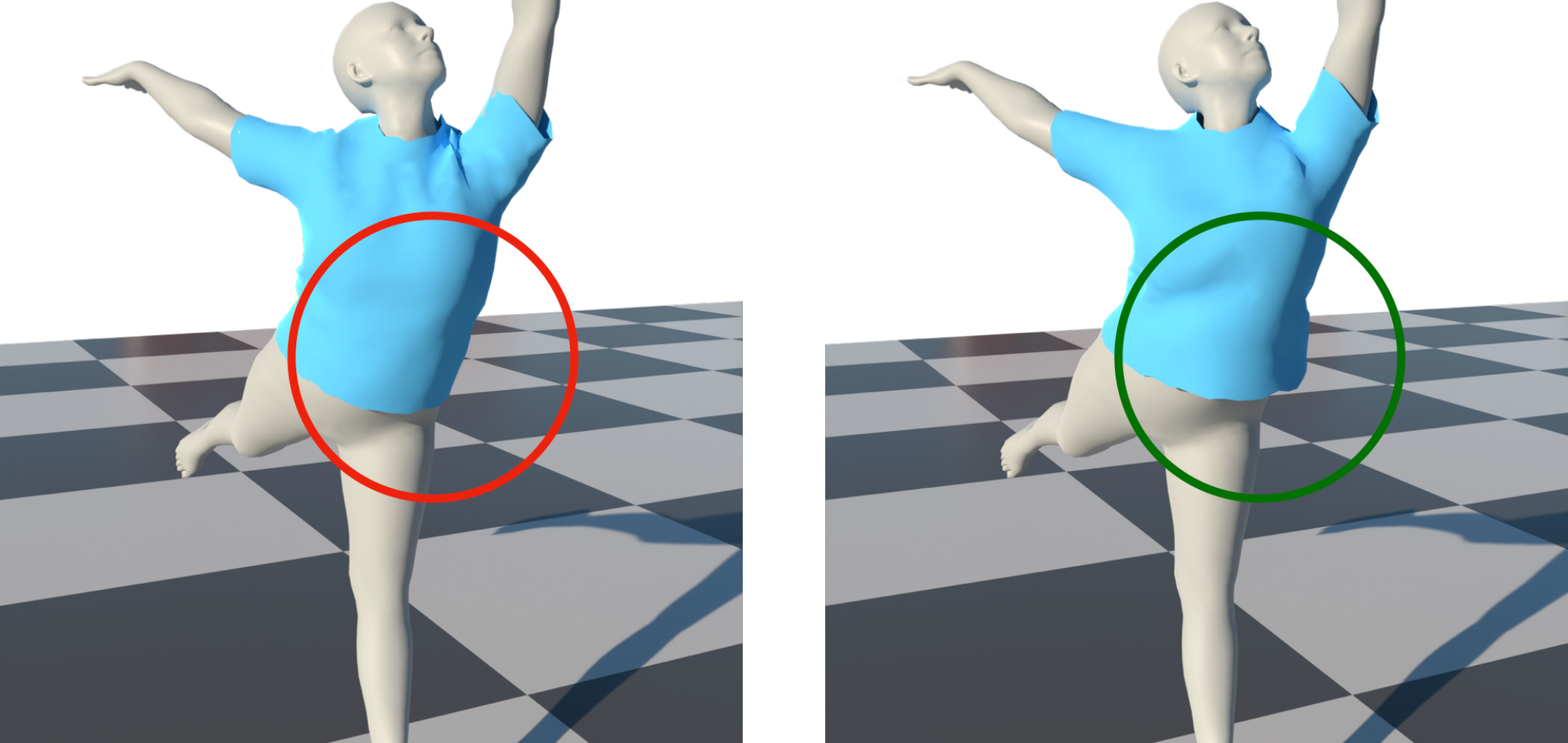}
      \put(12.5,  \pll){{\small SNUG~\cite{santesteban2022snug}}}
      \put(73,  \pll){{\small Ours}}
      \end{overpic}
    \caption{Comparison to SNUG. The global self-attention of our model allows the cloth to deform naturally in front of the torso, yielding better visual quality, while the baseline fails to generate the same effect.}
    \label{fig:comp-snug}
  \end{figure}
\textbf{Unsupervised methods.} We compare our method to the state-of-the-art of unsupervised learning method SNUG~\cite{santesteban2022snug}. We use the version of our method trained on the VTO dataset that this method is also trained on. Note that SNUG requires new training for every new garment while our method provides a general model. It can be seen in \Cref{fig:comp-snug} that our model is able to generate plausible results and provides better dynamics, while the garments are not seen during training. Please refer to the accompanying video for a complete result.    

\begin{figure}
    \centering
    \newcommand{\pll}{-5}
    \begin{overpic}[width=\columnwidth]{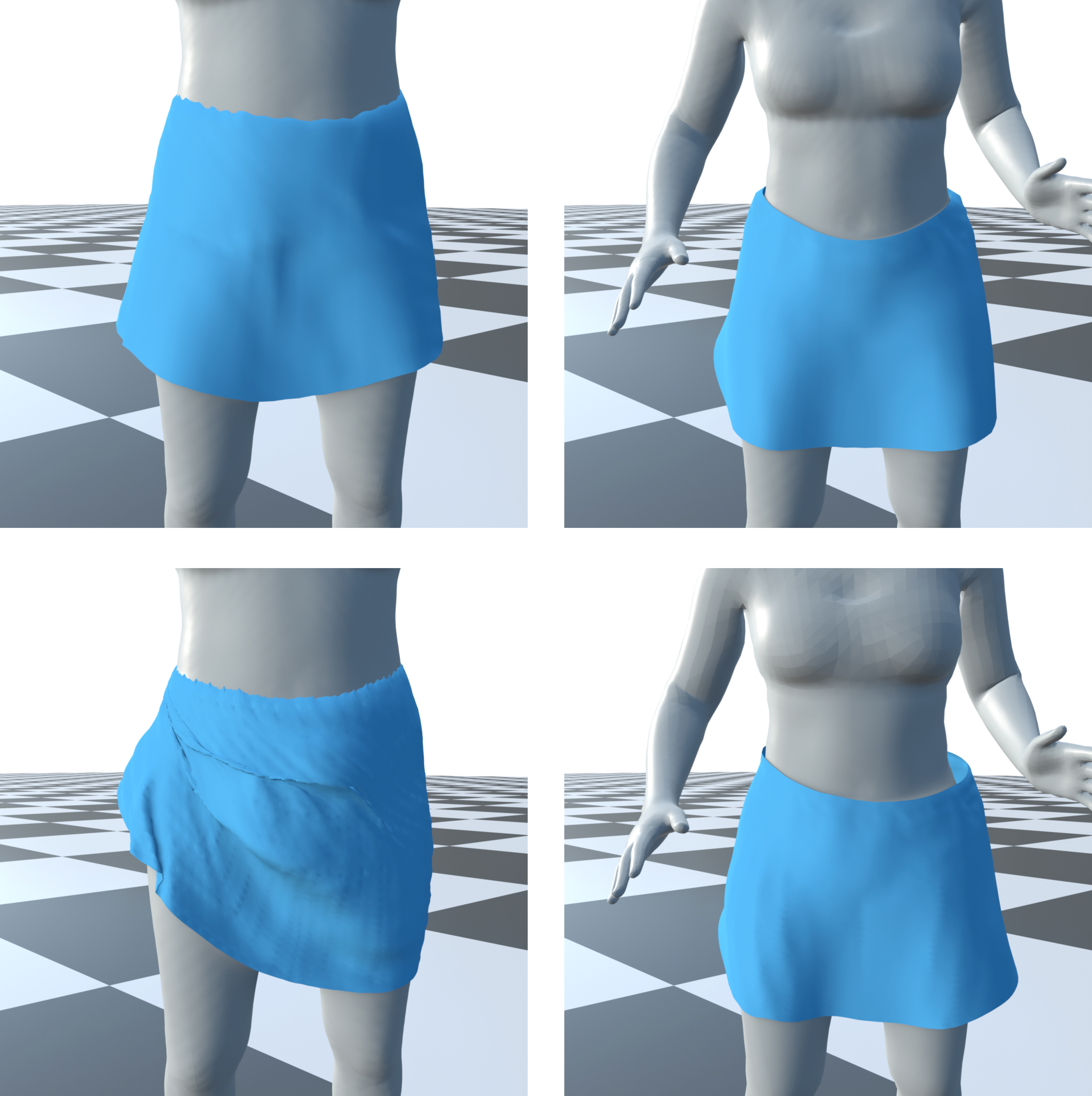}
      \put(12,  \pll){{\small HOOD~\cite{grigorev2023hood}}}
      \put(72,  \pll){{\small Ours}}
      \end{overpic}
    \caption{Comparison to HOOD. The meshes shown in the first and second rows contain 5,169 and 31,008 faces, respectively. It can be seen that our method produces consistent results across different resolutions, whereas HOOD exhibits unnatural dynamics and artifacts when processing high-resolution input.}
    \label{fig:comp-hood-2}
  \end{figure}
\textbf{Generalizable methods.} Finally, we compare our method to the concurrent works~\cite{tiwari2023garsim,grigorev2023hood} that tackle the generalization problem. As the code/date of GarSim~\cite{tiwari2023garsim} is not publicly available yet, we are limited to providing qualitative comparisons in the supplementary material.
To compensate for the limited receptive field of graph convolution, HOOD~\cite{grigorev2023hood} constructs a hierarchy of simplified meshes. As demonstrated in \Cref{fig:comp-hood-2}, when provided with high-resolution input, HOOD struggles to capture correct global dynamics due to the slow propagation of these dynamics and generates undesired artifacts caused by simplified graph construction. Please refer to the accompanying video for detailed results. 

\section{Conclusion}
In this paper, we introduce a learning-based generalizable solution for predicting garment dynamics with respect to an underlying body in motion. Previous learning-based approaches have been limited to garment-collider combinations present in the training dataset, often requiring refinement when applied to unseen cases. To address this limitation, we propose a deformation field-based garment representation combined with a transformer-based neural network. 
This combination enables us to handle different garment types and body models as colliders, contrary to existing methods, which are usually limited to a predefined parametric model (e.g., SMLP~\cite{SMPL2015}).
Additionally, we incorporate pairwise geodesic distances to weight the self-attention heads in our network, resulting in a manifold-aware transformer capable of capturing not only spatial but also topological correlations.

Our experiments demonstrate that our method can handle challenging scenarios involving complex garments and dynamic motions. We believe that our approach represents a novel direction toward modeling realistic and captivating clothing behavior for general garment dynamics and digital human modeling.

\textbf{Limitations and Future work.} 
To focus on our key insights, we have made several assumptions to simplify the problem setup. However, these assumptions also set limitations for our work and may stimulate potential future work.

\emph{Self-collision handling} We do not explicitly handle the garment-to-garment collisions. Thanks to our manifold-aware transformer, garment self-interpenetrations do not affect our auto-regressive workflow and are visually hard to observe according to our experiments. However, this could still be a problem for some downstream applications and may eventually lead to physically incorrect garment modeling. 

\emph{Material variation.} Our training dataset is produced by the same set of parameters, so the fabric material variation is not formulated into our current model. Allowing the users to control the material property can be a bonus feature for many scenarios~\cite{wang2018learning}. 

\emph{Fine details in predicted geometry.} Although our method can handle global dynamics well, the results lack fine details such as wrinkles when compared with state-of-the-art methods, as can be seen in the collar region in \Cref{fig:comp-snug}. We conjecture that this is due to the downsampling strategy and could be improved by reducing the downsampling rate and training over a longer time.

\emph{Unsupervised learning.} Last but not least, recent advances in unsupervised garment dynamics learning based on physical properties~\cite{bertiche2022neural, santesteban2022snug} show a promising direction with no data generation burdens. We believe having the terms from the physical constraints can potentially elaborate our approach in an unsupervised manner as well.

\section*{Acknowledgements}

We thank the anonymous reviewers for their valuable feedback. This work was supported in part by the ERC Consolidator Grant No. 101003104 (MYCLOTH), Adobe Research Internship Program, and Adobe University Gift Fund.

\printbibliography

@inproceedings{nealen2006physically,
  title={Physically based deformable models in computer graphics},
  author={Nealen, Andrew and M{\"u}ller, Matthias and Keiser, Richard and Boxerman, Eddy and Carlson, Mark},
  booktitle={Computer graphics forum},
  volume={25},
  number={4},
  pages={809--836},
  year={2006},
  organization={Wiley Online Library}
}

@article{muller2007position,
  title={Position based dynamics},
  author={M{\"u}ller, Matthias and Heidelberger, Bruno and Hennix, Marcus and Ratcliff, John},
  journal={Journal of Visual Communication and Image Representation},
  volume={18},
  number={2},
  pages={109--118},
  year={2007},
  publisher={Elsevier}
}

@article{muller2008hierarchical,
  title={Hierarchical position based dynamics},
  author={M{\"u}ller, Matthias},
  year={2008},
  publisher={The Eurographics Association}
}

@inproceedings{moore1988collision,
  title={Collision detection and response for computer animation},
  author={Moore, Matthew and Wilhelms, Jane},
  booktitle={Proceedings of the 15th annual conference on Computer graphics and interactive techniques},
  pages={289--298},
  year={1988}
}

@inproceedings{terzopoulos1987elastically,
  title={Elastically deformable models},
  author={Terzopoulos, Demetri and Platt, John and Barr, Alan and Fleischer, Kurt},
  booktitle={Proceedings of the 14th annual conference on Computer graphics and interactive techniques},
  pages={205--214},
  year={1987}
}

@article{bhat2003estimating,
  title={Estimating cloth simulation parameters from video},
  author={Bhat, Kiran S and Twigg, Christopher D and Hodgins, Jessica K and Khosla, Pradeep and Popovic, Zoran and Seitz, Steven M},
  year={2003},
  publisher={Carnegie Mellon University}
}

@inproceedings{miguel2012data,
  title={Data-driven estimation of cloth simulation models},
  author={Miguel, Eder and Bradley, Derek and Thomaszewski, Bernhard and Bickel, Bernd and Matusik, Wojciech and Otaduy, Miguel A and Marschner, Steve},
  booktitle={Computer Graphics Forum},
  volume={31},
  number={2pt2},
  pages={519--528},
  year={2012},
  organization={Wiley Online Library}
}

@article{liang2019differentiable,
  title={Differentiable cloth simulation for inverse problems},
  author={Liang, Junbang and Lin, Ming and Koltun, Vladlen},
  journal={Advances in Neural Information Processing Systems},
  volume={32},
  year={2019}
}

@article{li2022diffcloth,
  title={DiffCloth: Differentiable cloth simulation with dry frictional contact},
  author={Li, Yifei and Du, Tao and Wu, Kui and Xu, Jie and Matusik, Wojciech},
  journal={ACM Transactions on Graphics (TOG)},
  volume={42},
  number={1},
  pages={1--20},
  year={2022},
  publisher={ACM New York, NY}
}

@article{li2020codimensional,
  title={Codimensional incremental potential contact},
  author={Li, Minchen and Kaufman, Danny M and Jiang, Chenfanfu},
  journal={arXiv preprint arXiv:2012.04457},
  year={2020}
}

@incollection{harmon2009asynchronous,
  title={Asynchronous contact mechanics},
  author={Harmon, David and Vouga, Etienne and Smith, Breannan and Tamstorf, Rasmus and Grinspun, Eitan},
  booktitle={ACM SIGGRAPH 2009 papers},
  pages={1--12},
  year={2009}
}

@incollection{choi2005stable,
  title={Stable but responsive cloth},
  author={Choi, Kwang-Jin and Ko, Hyeong-Seok},
  booktitle={ACM SIGGRAPH 2005 Courses},
  pages={1--es},
  year={2005}
}

@article{volino2009simple,
  title={A simple approach to nonlinear tensile stiffness for accurate cloth simulation},
  author={Volino, Pascal and Magnenat-Thalmann, Nadia and Faure, Francois},
  journal={ACM Transactions on Graphics},
  volume={28},
  number={4},
  pages={Article--No},
  year={2009}
}

@inproceedings{lipman2004differential,
  title={Differential coordinates for interactive mesh editing},
  author={Lipman, Yaron and Sorkine, Olga and Cohen-Or, Daniel and Levin, David and Rossi, Christian and Seidel, Hans-Peter},
  booktitle={Proceedings Shape Modeling Applications, 2004.},
  pages={181--190},
  year={2004},
  organization={IEEE}
}

@article{narain2012adaptive,
  title={Adaptive anisotropic remeshing for cloth simulation},
  author={Narain, Rahul and Samii, Armin and O'brien, James F},
  journal={ACM transactions on graphics (TOG)},
  volume={31},
  number={6},
  pages={1--10},
  year={2012},
  publisher={ACM New York, NY, USA}
}

@article{yang2016detailed,
  title={Detailed garment recovery from a single-view image},
  author={Yang, Shan and Ambert, Tanya and Pan, Zherong and Wang, Ke and Yu, Licheng and Berg, Tamara and Lin, Ming C},
  journal={arXiv preprint arXiv:1608.01250},
  year={2016}
}

@inproceedings{zhang2021deep,
  title={Deep detail enhancement for any garment},
  author={Zhang, Meng and Wang, Tuanfeng and Ceylan, Duygu and Mitra, Niloy J},
  booktitle={Computer Graphics Forum},
  volume={40},
  number={2},
  pages={399--411},
  year={2021},
  organization={Wiley Online Library}
}

@article{wang2018learning,
  title={Learning a shared shape space for multimodal garment design},
  author={Wang, Tuanfeng Y and Ceylan, Duygu and Popovi{\'c}, Jovan and Mitra, Niloy J},
  journal={ACM Transactions on Graphics (TOG)},
  volume={37},
  number={6},
  pages={1--13},
  year={2018},
  publisher={ACM New York, NY, USA}
}

@article{bertiche2022neural,
  title={Neural Cloth Simulation},
  author={Bertiche, Hugo and Madadi, Meysam and Escalera, Sergio},
  journal={ACM Transactions on Graphics (TOG)},
  volume={41},
  number={6},
  pages={1--14},
  year={2022},
  publisher={ACM New York, NY, USA}
}

@article{guan2012drape,
  title={Drape: Dressing any person},
  author={Guan, Peng and Reiss, Loretta and Hirshberg, David A and Weiss, Alexander and Black, Michael J},
  journal={ACM Transactions on Graphics (ToG)},
  volume={31},
  number={4},
  pages={1--10},
  year={2012},
  publisher={ACM New York, NY, USA}
}

@inproceedings{patel2020tailornet,
  title={Tailornet: Predicting clothing in 3d as a function of human pose, shape and garment style},
  author={Patel, Chaitanya and Liao, Zhouyingcheng and Pons-Moll, Gerard},
  booktitle={Proceedings of the IEEE/CVF Conference on Computer Vision and Pattern Recognition},
  pages={7365--7375},
  year={2020}
}

@inproceedings{d2022n,
  title={N-Cloth: Predicting 3D Cloth Deformation with Mesh-Based Networks},
  author={D. Li, Y and Tang, Min and Yang, Yun and Huang, Zi and F. Tong, R and Yang, SC and Li, Yao and Manocha, Dinesh},
  booktitle={Computer Graphics Forum},
  volume={41},
  number={2},
  pages={547--558},
  year={2022},
  organization={Wiley Online Library}
}

@inproceedings{santesteban2021self,
  title={Self-supervised collision handling via generative 3d garment models for virtual try-on},
  author={Santesteban, Igor and Thuerey, Nils and Otaduy, Miguel A and Casas, Dan},
  booktitle={Proceedings of the IEEE/CVF Conference on Computer Vision and Pattern Recognition},
  pages={11763--11773},
  year={2021}
}

@article{bertiche2021pbns,
  title={PBNS: physically based neural simulation for unsupervised garment pose space deformation},
  author={Bertiche, Hugo and Madadi, Meysam and Escalera, Sergio},
  journal={ACM Transactions on Graphics (TOG)},
  volume={40},
  number={6},
  pages={1--14},
  year={2021},
  publisher={ACM New York, NY, USA}
}

@inproceedings{lahner2018deepwrinkles,
  title={Deepwrinkles: Accurate and realistic clothing modeling},
  author={Lahner, Zorah and Cremers, Daniel and Tung, Tony},
  booktitle={Proceedings of the European conference on computer vision (ECCV)},
  pages={667--684},
  year={2018}
}

@article{de2010stable,
  title={Stable spaces for real-time clothing},
  author={De Aguiar, Edilson and Sigal, Leonid and Treuille, Adrien and Hodgins, Jessica K},
  journal={ACM Transactions on Graphics (TOG)},
  volume={29},
  number={4},
  pages={1--9},
  year={2010},
  publisher={ACM New York, NY, USA}
}

@article{aigerman2022neural,
author = {Aigerman, Noam and Gupta, Kunal and Kim, Vladimir G. and Chaudhuri, Siddhartha and Saito, Jun and Groueix, Thibault},
title = {Neural Jacobian Fields: Learning Intrinsic Mappings of Arbitrary Meshes},
year = {2022},
issue_date = {July 2022},
publisher = {Association for Computing Machinery},
address = {New York, NY, USA},
volume = {41},
number = {4},
issn = {0730-0301},
url = {https://doi.org/10.1145/3528223.3530141},
doi = {10.1145/3528223.3530141},
journal = {ACM Trans. Graph.},
month = {jul},
articleno = {109},
numpages = {17}
}

@inproceedings{pan2022predicting,
  title={Predicting loose-fitting garment deformations using bone-driven motion networks},
  author={Pan, Xiaoyu and Mai, Jiaming and Jiang, Xinwei and Tang, Dongxue and Li, Jingxiang and Shao, Tianjia and Zhou, Kun and Jin, Xiaogang and Manocha, Dinesh},
  booktitle={ACM SIGGRAPH 2022 Conference Proceedings},
  pages={1--10},
  year={2022}
}

@inproceedings{holden2019subspace,
  title={Subspace neural physics: Fast data-driven interactive simulation},
  author={Holden, Daniel and Duong, Bang Chi and Datta, Sayantan and Nowrouzezahrai, Derek},
  booktitle={Proceedings of the 18th annual ACM SIGGRAPH/Eurographics Symposium on Computer Animation},
  pages={1--12},
  year={2019}
}

@article{habermann2021real,
  title={Real-time deep dynamic characters},
  author={Habermann, Marc and Liu, Lingjie and Xu, Weipeng and Zollhoefer, Michael and Pons-Moll, Gerard and Theobalt, Christian},
  journal={ACM Transactions on Graphics (TOG)},
  volume={40},
  number={4},
  pages={1--16},
  year={2021},
  publisher={ACM New York, NY, USA}
}

@article{vaswani2017attention,
  title={Attention is all you need},
  author={Vaswani, Ashish and Shazeer, Noam and Parmar, Niki and Uszkoreit, Jakob and Jones, Llion and Gomez, Aidan N and Kaiser, {\L}ukasz and Polosukhin, Illia},
  journal={Advances in neural information processing systems},
  volume={30},
  year={2017}
}

@article{guo2021pct,
  title={Pct: Point cloud transformer},
  author={Guo, Meng-Hao and Cai, Jun-Xiong and Liu, Zheng-Ning and Mu, Tai-Jiang and Martin, Ralph R and Hu, Shi-Min},
  journal={Computational Visual Media},
  volume={7},
  number={2},
  pages={187--199},
  year={2021},
  publisher={Springer}
}

@inproceedings{sorkine2004laplacian,
  title={Laplacian surface editing},
  author={Sorkine, Olga and Cohen-Or, Daniel and Lipman, Yaron and Alexa, Marc and R{\"o}ssl, Christian and Seidel, H-P},
  booktitle={Proceedings of the 2004 Eurographics/ACM SIGGRAPH symposium on Geometry processing},
  pages={175--184},
  year={2004}
}

@article{sumner2004deformation,
  title={Deformation transfer for triangle meshes},
  author={Sumner, Robert W and Popovi{\'c}, Jovan},
  journal={ACM Transactions on graphics (TOG)},
  volume={23},
  number={3},
  pages={399--405},
  year={2004},
  publisher={ACM New York, NY, USA}
}

@incollection{NEURIPS2019_9015,
title = {PyTorch: An Imperative Style, High-Performance Deep Learning Library},
author = {Paszke, Adam and Gross, Sam and Massa, Francisco and Lerer, Adam and Bradbury, James and Chanan, Gregory and Killeen, Trevor and Lin, Zeming and Gimelshein, Natalia and Antiga, Luca and Desmaison, Alban and Kopf, Andreas and Yang, Edward and DeVito, Zachary and Raison, Martin and Tejani, Alykhan and Chilamkurthy, Sasank and Steiner, Benoit and Fang, Lu and Bai, Junjie and Chintala, Soumith},
booktitle = {Advances in Neural Information Processing Systems 32},
editor = {H. Wallach and H. Larochelle and A. Beygelzimer and F. d\textquotesingle Alch\'{e}-Buc and E. Fox and R. Garnett},
pages = {8024--8035},
year = {2019},
publisher = {Curran Associates, Inc.},
url = {http://papers.neurips.cc/paper/9015-pytorch-an-imperative-style-high-performance-deep-learning-library.pdf}
}

@article{kingma2014adam,
  title={Adam: A method for stochastic optimization},
  author={Kingma, Diederik P and Ba, Jimmy},
  journal={arXiv preprint arXiv:1412.6980},
  year={2014}
}

@article{SMPL2015,
  author = {Loper, Matthew and Mahmood, Naureen and Romero, Javier and Pons-Moll, Gerard and Black, Michael J.},
  title = {{SMPL}: A Skinned Multi-Person Linear Model},
  journal = {ACM Trans. Graphics (Proc. SIGGRAPH Asia)},
  month = oct,
  number = {6},
  pages = {248:1--248:16},
  publisher = {ACM},
  volume = {34},
  year = {2015}
}

@inproceedings{tiwari2023garsim,
  title={GarSim: Particle Based Neural Garment Simulator},
  author={Tiwari, Lokender and Bhowmick, Brojeshwar},
  booktitle={Proceedings of the IEEE/CVF Winter Conference on Applications of Computer Vision},
  pages={4472--4481},
  year={2023}
}

@article{dwivedi2020generalization,
  title={A generalization of transformer networks to graphs},
  author={Dwivedi, Vijay Prakash and Bresson, Xavier},
  journal={arXiv preprint arXiv:2012.09699},
  year={2020}
}

@article{pfaff2020learning,
  title={Learning mesh-based simulation with graph networks},
  author={Pfaff, Tobias and Fortunato, Meire and Sanchez-Gonzalez, Alvaro and Battaglia, Peter W},
  journal={arXiv preprint arXiv:2010.03409},
  year={2020}
}

@inproceedings{grigorev2023hood,
  author = {Grigorev, Artur and Thomaszewski, Bernhard and Black, Michael J and Hilliges, Otmar}, 
  title = {{HOOD}: Hierarchical Graphs for Generalized Modelling of Clothing Dynamics}, 
  journal = {Computer Vision and Pattern Recognition (CVPR)},
  year = {2023},
}

@article{de2022drapenet,
  title={DrapeNet: Generating Garments and Draping them with Self-Supervision},
  author={De Luigi, Luca and Li, Ren and Guillard, Beno{\^i}t and Salzmann, Mathieu and Fua, Pascal},
  journal={arXiv preprint arXiv:2211.11277},
  year={2022}
}

@inproceedings{chandran2022shape,
  title={Shape Transformers: Topology-Independent 3D Shape Models Using Transformers},
  author={Chandran, Prashanth and Zoss, Gaspard and Gross, Markus and Gotardo, Paulo and Bradley, Derek},
  booktitle={Computer Graphics Forum},
  volume={41},
  number={2},
  pages={195--207},
  year={2022},
  organization={Wiley Online Library}
}

@inproceedings{bertiche2020cloth3d,
  title={CLOTH3D: clothed 3d humans},
  author={Bertiche, Hugo and Madadi, Meysam and Escalera, Sergio},
  booktitle={Computer Vision--ECCV 2020: 16th European Conference, Glasgow, UK, August 23--28, 2020, Proceedings, Part XX 16},
  pages={344--359},
  year={2020},
  organization={Springer}
}

@inproceedings{santesteban2022snug,
  title={Snug: Self-supervised neural dynamic garments},
  author={Santesteban, Igor and Otaduy, Miguel A and Casas, Dan},
  booktitle={Proceedings of the IEEE/CVF Conference on Computer Vision and Pattern Recognition},
  pages={8140--8150},
  year={2022}
}

@article{wu2021density,
  title={Density-aware chamfer distance as a comprehensive metric for point cloud completion},
  author={Wu, Tong and Pan, Liang and Zhang, Junzhe and Wang, Tai and Liu, Ziwei and Lin, Dahua},
  journal={arXiv preprint arXiv:2111.12702},
  year={2021}
}

@url{CMUmocap,
 Author = {{CMU}},
 Year = 2019,
 Month = {May},
 Lastchecked = {May 2019},
 Title = {{CMU Graphics Lab Motion Capture Database}},
 Url = {http://mocap.cs.cmu.edu/}
}

@inproceedings{jin2020pixel,
  title={A pixel-based framework for data-driven clothing},
  author={Jin, Ning and Zhu, Yilin and Geng, Zhenglin and Fedkiw, Ronald},
  booktitle={Computer Graphics Forum},
  volume={39},
  number={8},
  pages={135--144},
  year={2020},
  organization={Wiley Online Library}
}

@inproceedings{gundogdu2019garnet,
  title={Garnet: A two-stream network for fast and accurate 3d cloth draping},
  author={Gundogdu, Erhan and Constantin, Victor and Seifoddini, Amrollah and Dang, Minh and Salzmann, Mathieu and Fua, Pascal},
  booktitle={Proceedings of the IEEE/CVF International Conference on Computer Vision},
  pages={8739--8748},
  year={2019}
}

@article{zhang2022motion,
  title={Motion guided deep dynamic 3d garments},
  author={Zhang, Meng and Ceylan, Duygu and Mitra, Niloy J},
  journal={ACM Transactions on Graphics (TOG)},
  volume={41},
  number={6},
  pages={1--12},
  year={2022},
  publisher={ACM New York, NY, USA}
}

@inproceedings{ying2023adaptive,
  title={Adaptive Local Basis Functions for Shape Completion},
  author={Ying, Hui and Shao, Tianjia and Wang, He and Yang, Yin and Zhou, Kun},
  booktitle={ACM SIGGRAPH 2023 Conference Proceedings},
  pages={1--11},
  year={2023}
}

@article{luo2018nnwarp,
  title={NNWarp: Neural network-based nonlinear deformation},
  author={Luo, Ran and Shao, Tianjia and Wang, Huamin and Xu, Weiwei and Chen, Xiang and Zhou, Kun and Yang, Yin},
  journal={IEEE transactions on visualization and computer graphics},
  volume={26},
  number={4},
  pages={1745--1759},
  year={2018},
  publisher={IEEE}
}

@article{cholmod,
author = {Chen, Yanqing and Davis, Timothy A. and Hager, William W. and Rajamanickam, Sivasankaran},
title = {Algorithm 887: CHOLMOD, Supernodal Sparse Cholesky Factorization and Update/Downdate},
year = {2008},
issue_date = {October 2008},
publisher = {Association for Computing Machinery},
address = {New York, NY, USA},
volume = {35},
number = {3},
issn = {0098-3500},
url = {https://doi.org/10.1145/1391989.1391995},
doi = {10.1145/1391989.1391995},
journal = {ACM Trans. Math. Softw.},
month = {oct},
articleno = {22},
numpages = {14}
}

@article{naumov2011parallel,
  title={Parallel solution of sparse triangular linear systems in the preconditioned iterative methods on the GPU},
  author={Naumov, Maxim},
  journal={NVIDIA Corp., Westford, MA, USA, Tech. Rep. NVR-2011},
  volume={1},
  year={2011}
}

@article{nicolet2021large,
  title={Large steps in inverse rendering of geometry},
  author={Nicolet, Baptiste and Jacobson, Alec and Jakob, Wenzel},
  journal={ACM Transactions on Graphics (TOG)},
  volume={40},
  number={6},
  pages={1--13},
  year={2021},
  publisher={ACM New York, NY, USA}
}

@article{jacobson2013robust,
  title={Robust inside-outside segmentation using generalized winding numbers},
  author={Jacobson, Alec and Kavan, Ladislav and Sorkine-Hornung, Olga},
  journal={ACM Transactions on Graphics (TOG)},
  volume={32},
  number={4},
  pages={1--12},
  year={2013},
  publisher={ACM New York, NY, USA}
}

\end{document}

\title{Supplementary material: \\
Neural Garment Dynamics via Manifold-Aware Transformers
}

\author[Li, P.\ et al.] 
{\parbox{\textwidth}{\centering Peizhuo Li$^{1}$\orcid{0000-0001-9309-9967}, Tuanfeng Y. Wang$^{2}$\orcid{0000-0002-8180-4988}, Timur Levent Kesdogan$^{1}$\orcid{0009-0006-5839-9677}, Duygu Ceylan$^{2}$\orcid{0000-0001-6530-4556}, Olga Sorkine-Hornung$^{1}$\orcid{0000-0002-8089-3974}
        }
        \\
% For Computer Graphics Forum: Please use the abbreviation of your first name.
{\parbox{\textwidth}{\centering $^1$ETH Zurich, Switzerland $\quad \quad \quad$
         $^2$Adobe Research, United Kingdom
       }
}
}

\newif\ifdraft
\drafttrue
% \draftfalse

\maketitle

\section{Deformation Gradients}

In this section, we review the preliminaries for the deformation gradients and Poisson equation.

\subsection{Deformation gradients field}

Given a discretized garment mesh, assumed to be a 2-manifold triangular mesh with vertices position $\Vrest$ at rest status and triangulation $\bbt$. For a triangle $i$ associated with vertices $v_j, v_k, v_l$ (counter-clockwise), its local coordinate $Q_i$ as a $3 \times 3$ matrix is defined as $\left[ v_k - v_j, v_l - v_j, \bbn_i \right]$ and $\bbn_i$ is the unit outward normal of triangle $i$. The deformation gradient of triangle $i$ at frame $t$ with vertices position $\bbv^t$ w.r.t. to the rest status can then be defined as $\Phi_i^t = Q_i^t Q_i^{-\mathrm{rest}}$, where $Q_i^{-\mathrm{rest}}$ refers to matrix inversion of $Q_i^{\mathrm{rest}}$, the local coordinate of triangle $j$ at rest status. We denote the deformation gradient field at frame $t$ as $\Phi^t := \{\Phi_i^t\}_{i \in \bbt}$. 
We choose deformation gradients because they fully capture the local deformation of the garment and are invariant to the global translation and rotation.

\subsection{Relative deformation gradient} Although the deformation gradients field 
provides a triangulation-agnostic representation, directly predicting the deformation gradient of frame $i$ based on features of previous frames is not the most efficient way. Instead, the residual of the deformation gradient field between two frames, namely the relative deformation gradients $\Psi^t = \Phi^t \Phi^{-(t-1)}$, where $\Phi^{-(t-1)}$ is the element-wise matrix inverse of $\Phi^{t-1}$, is more suitable for learning the dynamics of the garment deformation and frees the network from the effort of remembering the absolute deformation from previous frames.

\subsection{Poisson equation} Given an arbitrary deformation gradient field $\Phi^t$, 
we can reconstruct the deformed mesh by solving a Poisson equation:

\begin{equation}
    \bbv^* = \argmin_{\bbv} \sum_{i \in \bbt} s_i \left\|\Phi_i(\bbv) - \Phi_i^t \right\|_F^2,
    \label{eq:poisson}
\end{equation}
where $s_i$ is the area of triangle $i$ and $\Phi_i(\bbv)$ is the deformation gradient of triangle $i$ given vertex positions $\bbv$. It is a sparse linear system w.r.t. to $\bbv$ and can be efficiently solved with the Laplace-Beltrami operator~\cite{sumner2004deformation, sorkine2004laplacian}.

\section{Collision refinement}

\begin{figure}
    \centering
    \newcommand{\pll}{-5}
    \begin{overpic}[width=\columnwidth]{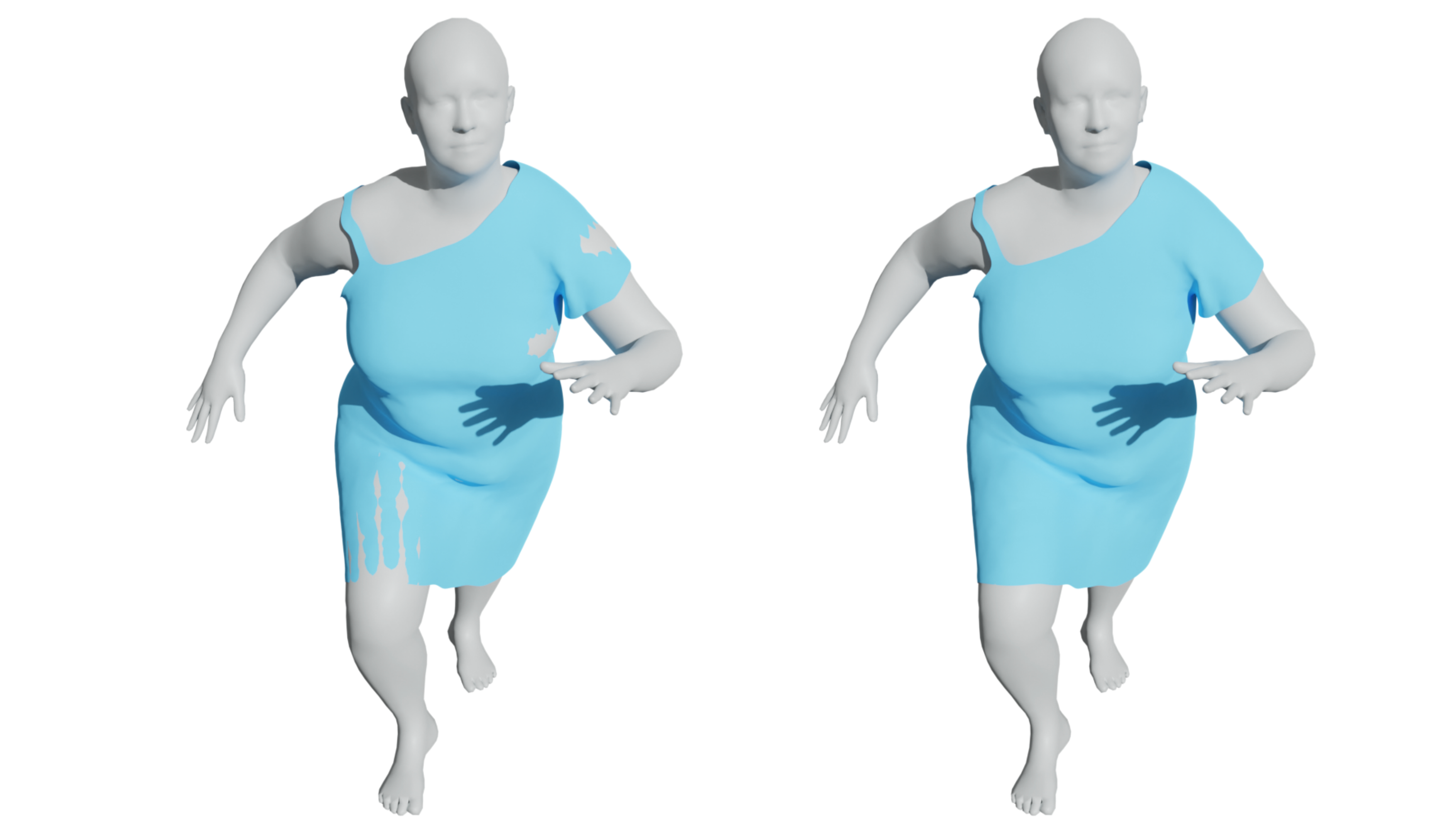}
      \put(26,  \pll){{\small Before}}
      \put(70,  \pll){{\small After}}
      \end{overpic}
    \caption{Collision refinement. Our post-process refines the collision between garment and body in raw prediction by minimizing the proposed energy.}
    \label{fig:collision}
  \end{figure}

We use a post-process adopted from DRAPE~\cite{guan2012drape} for collision refinement. Assuming the predicted cloth's vertex position is $\tilde\bbv$ and the body's vertex position is $\bbu$, we solve for a new vertex position $\bbv$ to minimize the following energy functions:
\begin{equation}
    E_{\text{collision}} = \sum_{(i, j) \in C} \| \epsilon + \vec n_j \cdot (\bbv_i - \bbu_j) \|_2^2,
    \label{eq:e_collision}
\end{equation}
where $C$ is the set containing the paired indices of the cloth vertex $i$ in collision with the body and its nearest body vertex $j$, and $\vec n_j$ is the outward normal of the body vertex $j$. This energy pushes the vertices inside the body away from the body surface. Besides, we also want to keep the local geometry of the cloth unchanged. We thus also include the following Laplacian term:
\begin{equation}
    E_{\text{lap}} = \| \Delta \bbv - \Delta \tilde\bbv\|_2^2.
    \label{eq:e_lap}
\end{equation}
To make this system determined, we add the following regularization term:
\begin{equation}
    E_{\text{reg}} = \| \bbv - \tilde\bbv \|_2^2.
    \label{eq:e_reg}
\end{equation}
The overall energy function is thus:
\begin{equation}
    E = E_{\text{collision}} + \lambda_{\text{lap}} E_{\text{lap}} + \lambda_{\text{reg}} E_{\text{reg}},
    \label{eq:e_overall}
\end{equation}
and we use $\lambda_{\text{lap}} = 0.5$ and $\lambda_{\text{reg}} = 1\times 10^{-3}$ in our experiments. It is a sparse least-square problem and can be efficiently solved with Cholesky decomposition. We show an example of collision refinement in \Cref{fig:collision}.

\section{Qualitative Comparisons}

\begin{figure}
    \centering
    \newcommand{\pll}{-5}
    \begin{overpic}[width=\columnwidth]{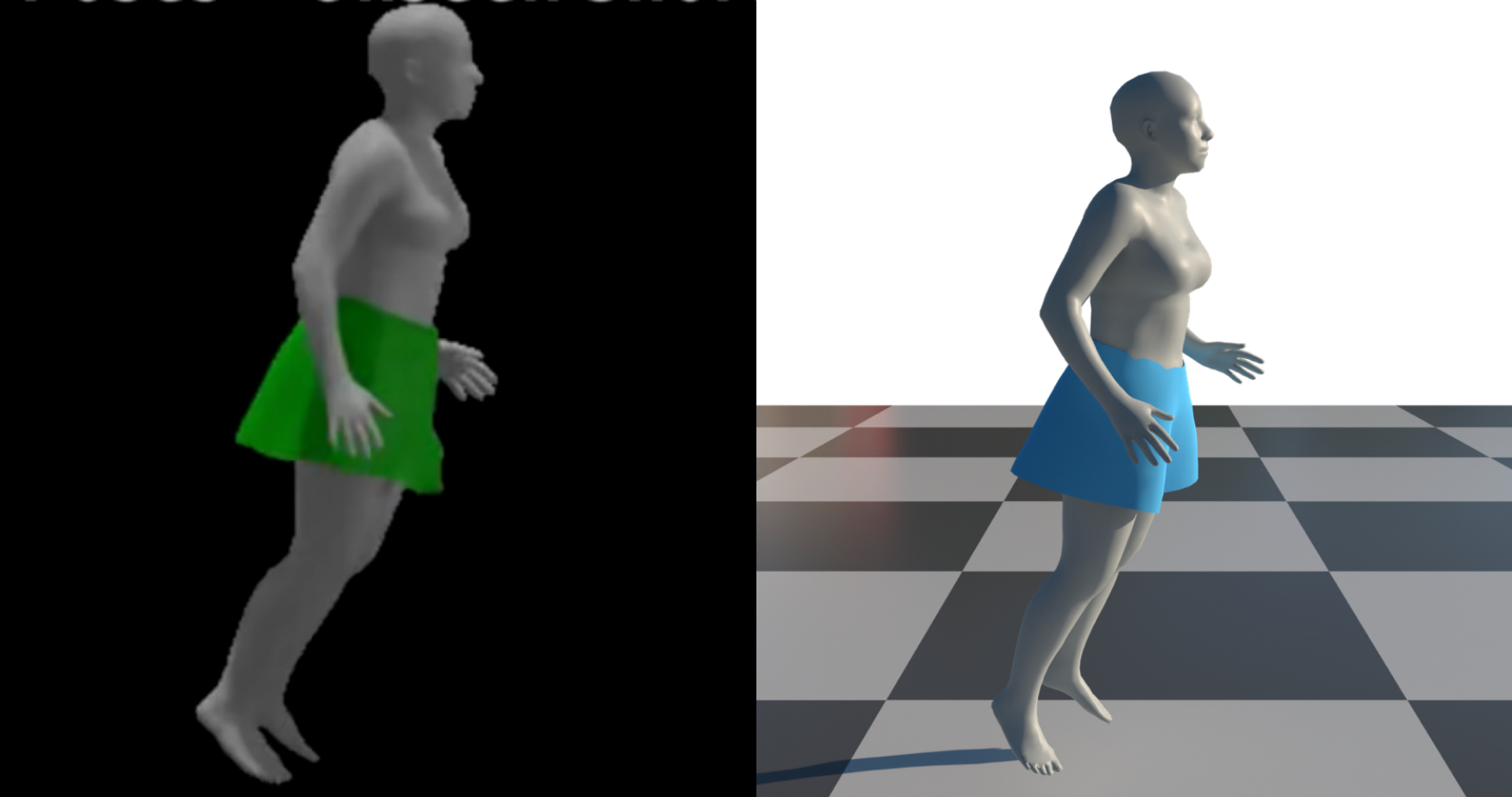}
      \put(15,  \pll){{\small GarSim~\cite{tiwari2023garsim}}}
      \put(72,  \pll){{\small Ours}}
      \end{overpic}
      \caption{Qualitative comparison to GarSim.}
    \label{fig:comp-garsim}
  \end{figure}

In this section, we qualitatively compare our results to the results of GarSim~\cite{tiwari2023garsim}. As can be seen in \Cref{fig:comp-garsim}, our model is able to create more dynamics thanks to the global awareness introduced by our manifold-aware transformer architecture. 

\section{Network architecture}

In this section, we describe the detailed network architecture of our framework. 

We use an encoder-only transformer~\cite{vaswani2017attention} architecture. Our model contains 8 layers of transformer block. Each transformer block contains a multi-head self-attention layer and a feed-forward layer. The embedding dimension and the feed-forward layer dimension are set to 512. The number of heads is set to 8. We use a dropout rate of 0.1. 

For the input of the network, we gather the feature extracted from the past 10 frames, namely $\nhist = 10$. The range of manifold-aware self-attention is controlled by $\pgeo = 20$. We choose to use $\nconn = 2$ manifold-aware self-attention heads out of 8 heads. During training time, we split the input geometry into $n_s = 4$ disjoint subsets in the first 100 epochs for efficiency consideration. We then stop the splitting (i.e. $n_s = 1$) for the remaining epochs to enable the network to learn fine details of garments. We train the network with a batch size of 16.

We use hyper-parameters $\lambda_{\text{sv}}$ and $\lambda_{\text{vel}}$ to balance our global loss term. We find that an equal weighting of the singular value loss $\Loss_{\text{sv}}$ and a higher weighting of the global velocity loss $\Loss_{\text{vel}}$ delivers optimal results. Thus, we use  $\lambda_{\text{sv}} = 1$ and $\lambda_{\text{vel}} = 3$ for all our experiments.

\printbibliography